\documentclass{article}
\usepackage{epsfig,endnotes}
\usepackage[english]{babel}
\usepackage[utf8]{inputenc}
\usepackage[T1]{fontenc} 
\usepackage{authblk}
\usepackage{listings}
\usepackage{booktabs}
\usepackage{pifont}
\usepackage[x11names,dvipsnames,table]{xcolor}
\usepackage{rotating}
\usepackage[inline]{enumitem}
\usepackage[a4paper, margin=1.2in]{geometry}
\usepackage{adjustbox} 
\usepackage{hyperref}
\usepackage{tcolorbox}
\usepackage{amsmath}
\usepackage{amsthm}
\newtheorem{definition}{Definition}
\newtheorem{theorem}{Theorem}
\newtheorem{corollary}{Corollary}

\usepackage{pgfplots}
\usepackage{xurl}
\pgfplotsset{compat=1.18}

\hypersetup{
    colorlinks=true,
    linkcolor=blue,
    citecolor=blue,
    filecolor=blue,      
    urlcolor=blue,
    pdfpagemode=FullScreen,
}

\lstdefinelanguage{json}{
    numbers=left,
    numberstyle=\small,
    rulecolor=\color{black},
    backgroundcolor=\color{col1},
    showspaces=false,
    showtabs=false,
    breaklines=true,
    postbreak=\raisebox{0ex}[0ex][0ex]{\ensuremath{\color{gray}\hookrightarrow\space}},
    breakatwhitespace=true,
    basicstyle=\ttfamily\small,
    upquote=true,
    morestring=[b]",
    stringstyle=\color{col7},
    literate=
     *{0}{{{\color{col5}0}}}{1}
      {1}{{{\color{col5}1}}}{1}
      {2}{{{\color{col5}2}}}{1}
      {3}{{{\color{col5}3}}}{1}
      {4}{{{\color{col5}4}}}{1}
      {5}{{{\color{col5}5}}}{1}
      {6}{{{\color{col5}6}}}{1}
      {7}{{{\color{col5}7}}}{1}
      {8}{{{\color{col5}8}}}{1}
      {9}{{{\color{col5}9}}}{1}
      {\{}{{{\color{col5}{\{}}}}{1}
      {\}}{{{\color{col5}{\}}}}}{1}
      {[}{{{\color{col5}{[}}}}{1}
      {]}{{{\color{col5}{]}}}}{1},
}

\definecolor{col1}{HTML}{fafafa}
\definecolor{col2}{HTML}{e4e5f1}
\definecolor{col3}{HTML}{d2d3db}
\definecolor{col4}{HTML}{9394a5}
\definecolor{col5}{HTML}{484b6a}
\definecolor{col6}{HTML}{eacdcd}
\definecolor{col7}{HTML}{e26e73}
\usepackage{amssymb}
\newcommand{\xmark}{\ding{55}}%
\newcommand{\cmark}{\ding{51}}%
\newcommand{\mycheck}{\textcolor{ForestGreen}{\cmark}}
\newcommand{\myxmark}{\textcolor{red}{\xmark}}
\newcommand{\myunknown}{\textcolor{teal}{\ding{108}}}
\PassOptionsToPackage{hyphens}{url}

\usepackage{tikz}
\usetikzlibrary{arrows.meta, positioning, calc}
\usetikzlibrary{shapes.symbols}

\tikzset{
  tcaStage/.style={
    draw, rounded corners=2pt, align=center,
    inner sep=4pt, font=\scriptsize, minimum height=8mm,
    text width=24mm
  },
  tcaNoteL/.style={
    draw, rounded corners=2pt, align=left,
    inner sep=4pt, font=\scriptsize,
    text width=60mm
  },
  tcaNoteR/.style={
    draw, rounded corners=2pt, align=left,
    inner sep=4pt, font=\scriptsize,
    text width=57mm
  },
  tcaArrow/.style={-Latex, line width=0.6pt},
  tcaCallout/.style={-Latex, line width=0.6pt},
}
\begin{document}

\title{Evading and crashing anti-malware solutions via data collection overloading during analysis serialization}

\author[1]{Evgenios Gkritsis}
\author[2,3]{Constantinos Patsakis}
\author[1]{George Stergiopoulos}

\affil[1]{Department of Informatics, Athens University of Economics and Business, Greece}
\affil[2]{Department of Informatics, University of Piraeus, 80 Karaoli \& Dimitriou str., 18534 Piraeus, Greece}
\affil[3]{Information Management Systems Institute of Athena Research Centre, Greece}

\maketitle

\begin{abstract}
  Malware analysis systems, including dynamic-analysis sandboxes and digital forensics and incident response (DFIR) platforms, rely on telemetry pipelines comprising collection agents, serializers, and database backends to capture and present program behavior to analysts. We show that these data-handling components constitute an exploitable attack surface that can lead to denial-of-analysis (DoA) states without disabling sensors or requiring elevated privileges. We present \textit{Telemetry Complexity Attacks} (TCAs), a new class of vulnerabilities that exploit mismatches between unbounded collection mechanisms and bounded processing capabilities. Our method recursively spawns child processes to generate deeply nested and oversized objects that stress serialization and storage boundaries, as well as visualization layers, e.g., JSON/BSON depth and size limits. Depending on the product, this leads to truncated or missing behavioral reports, rejected database inserts, serializer recursion and size errors, and unresponsive dashboards, with some cases also exhibiting normal malicious execution that was not recorded or presented to analysts. We evaluate our technique against 18 commercial and open-source malware analysis platforms and endpoint detection and response (EDR) solutions. Seven products fail at different stages of the telemetry pipeline; two CVE identifiers have been assigned (CVE-61301 and CVE-61303); one more is pending; one has been assigned to an underlying library, and others have issued patches or configuration changes. We discuss root causes and propose mitigation strategies to prevent DoA attacks triggered by adversarial telemetry.
  
\end{abstract}

\section{Introduction}

Security reports estimate that hundreds of thousands of new malicious programs emerge daily, contributing to a corpus exceeding one billion malware samples \cite{AVTEST2023}. The economic impact is similarly stark: IBM reports an average data-breach cost of \$4.4m, with a substantial share attributable to malware-driven incidents~\cite{ibm}. Ransomwhere \cite{cable_2024_13999026}, which catalogs cryptocurrency addresses used by ransomware groups, indicates that total tracked ransomware payments surpass \$1bn. CrowdStrike further attributes a 22\% reduction in average breakout time to infostealers \cite{crowdstrike}. Together, these figures underscore the financial consequences of modern malware, even before accounting for activity by advanced persistent threat (APT) groups and other threat actors.

Defenders often rely on telemetry-centric pipelines to detect malware. Endpoint detection and response (EDR) agents collect behavioral signals (e.g., process, network, file, and registry activity) for detection and investigation, while dynamic-analysis sandboxes detonate artifacts in isolated environments to observe behavior and generate reports with behavioral indicators \cite{MSDefenderEDROverview,CrowdStrikeEDRDefinition,SentinelOneBehavioralAI,PANWAdvancedWildFireEnv,PANWCortexXDRWildFireReport,CiscoMerakiThreatGridIntegration,CiscoThreatGridDatasheet,FortinetSandboxingEbook}. These systems typically combine OS-level event providers (e.g., process creation monitoring via API hooking or Event Tracing for Windows (ETW)/Sysmon instrumentation), user-space collectors, sandbox execution, and backend aggregation. For interoperability and storage, telemetry is commonly serialized into structured logs (often JSON) as it flows through the pipeline; for instance, Wazuh serializes endpoint events as JSON records (e.g., \texttt{alerts.json}) prior to indexing \cite{WazuhDocs2023}.

Such pipelines implicitly assume end-to-end resilience: once malicious activity executes in a monitored environment, its corresponding events should appear in the final report. In practice, malware undermines this assumption via anti-analysis techniques that detect sandboxed or instrumented settings and then alter, delay, or suppress behavior. Common strategies include checking for VM artifacts, timing and resource probes, and user-interaction tests, and have grown increasingly sophisticated \cite{Afianian2018}. These techniques primarily aim to evade observation by minimizing telemetry generation, often by making malware appear benign and exploiting defenders' dependence on novelty-based detection (e.g., new signatures or behavior models).

Our work departs from this paradigm. Telemetry pipelines strive to capture rich detail (event logs, process trees, memory snapshots), yet the components that serialize, store, and render this information have practical limits, including maximum JSON nesting depth, document size caps, database insertion constraints, and user interface (UI) rendering capacity. Here, we identify a critical design assumption: telemetry is treated as benign input to be processed, rather than as an attack surface. Defensive tools scrutinize program behavior but largely trust the infrastructure that records and transmits that behavior. Unlike traditional evasive malware that hides its actions, our attacks do not conceal malicious behavior; instead, they overload the pipeline's ability to represent it. The result is telemetry that is sufficiently malformed or voluminous that the security product fails to record it correctly or present it to an analyst, making the activity effectively invisible not by avoiding the observer, but by breaking it. We show that an attacker can exploit the telemetry pipeline itself using perfectly valid behaviors to stress data handling, achieving detection bypass without introducing new evasion of the underlying monitoring mechanisms.

\noindent\textbf{Relation to prior work.}
Our attacks are conceptually related to classic structured-data and resource-exhaustion techniques, including the ``billion laughs'' XML entity expansion and fork bombs. The ``billion laughs'' attack was initially targeted at XML parsers~\cite{198467}, but analogous attacks apply to other parsers, including YAML \cite{8887385} and SAML \cite{180226}. Fork bombs exhaust resource limits by rapidly spawning processes~\cite{berlot2008dealing}. Our approach lies between these two lines of work: like billion-laughs, we exploit pathological data structures; like fork bombs, we exploit process creation to exhaust downstream resources. However, we pursue different system- and architecture-specific goals and guarantee payload execution. In this regard, our work is closely related to Filic et al.~\cite{DBLP:conf/codaspy/FilicHMPU25}, who exploit implementations of well-known probabilistic data structures across multiple solutions. Contrary to arbitrary spawning in fork bombs, we have guarantees that the payload is executed, without triggering infrastructure failures that would alert the victim. The analyst receives either a `benign' verdict or an incomplete report lacking evidence of malicious behavior.

Our focus is denial-of-service against anti-malware and malware sandbox pipelines, but differs from prior approaches by targeting weaknesses in data structures and telemetry-handling paths rather than containment escape or instrumentation evasion. Instead of hiding or escaping analysis, we weaponize the data-handling components of monitoring systems to induce denial-of-service, analysis failure, or silent visibility loss. We show that such structures and failure modes are widely present in deployed anti-malware solutions and that this attack surface remains underexplored, spanning evasion, regex-based availability, and related attacks.

\noindent\textbf{Contributions.}
We make the following contributions. First, we introduce a subclass of exploits against malware analysis pipelines driven by a fundamental mismatch between unbounded data collection and bounded data processing and formalize \emph{Telemetry Complexity Attacks} (TCAs) as a subset of resource-exhaustion attacks with telemetry-specific failure modes. 

Secondly, we propose a process-spawning methodology that generates high-volume telemetry and targets the post-collection pipeline (serializers, databases, renderers), stressing serialization and storage boundaries and inducing analysis failure or denial-of-service in downstream components.

Finally, we evaluate the technique against multiple anti-malware solutions. We demonstrate that this method affects multiple real-world anti-malware products and services. Across deployments, we observe: \begin{enumerate*}[label=(\roman*)]
    \item missing or partial behavioral reports,
    \item rejected database inserts due to size or depth limits,
    \item serializer crashes from recursion or memory exhaustion, and
    \item stalled or unresponsive dashboards.
\end{enumerate*}
These effects require no elevated privileges and do not tamper with hooks, sensors, or agents; they arise solely from adversarially structured telemetry that is valid but pathological. Our evaluation resulted in vendor patches and two assigned CVEs, with one more pending, and one CVE affecting a dependency used by some of these systems. We provide an analysis and a binary proof-of-concept that reaches dynamic analysis engines (sandboxes and EDR telemetry pipelines) while avoiding static classification as malware. 

Our results show that the telemetry pipeline itself constitutes a critical attack surface in systems designed for behavioral observation and forensic analysis. To our knowledge, this is the first work to systematically demonstrate, across multiple deployed security products, that adversarial but valid telemetry structures can induce analyst-visible denial-of-analysis outcomes through serialization, storage, and rendering boundaries, even when monitoring sensors remain intact. Accordingly, we identify the telemetry pipeline as a cross-product attack surface, characterize failure modes across diverse architectures, and validate real-world impact via coordinated disclosures and assigned CVEs.

\noindent\textbf{Responsible disclosure.}
We coordinated responsible disclosure with affected vendors, providing at least 90 days for them to patch their solutions, leading to patches and configuration hardening across multiple EDR and sandbox products. At the time of writing, three issues have been assigned CVE IDs, namely CVE-61301 (CAPEv2), CVE-61303 (Triage), and CVE-67221 
(orjson serialization library), with one additional CVE pending for Cuckoo. Proof-of-concept artifacts for the three CVEs are available on GitHub~\cite{61301_poc,61303_poc,67221_poc}.

Overall, TCAs offer a new evasion perspective: by leveraging data serialization and pipeline processing limits, attackers can induce silent analysis failures without defeating the monitoring mechanisms themselves.

\section{Concept and methodology}
\label{sec:methodology}

Our objective is to stress the telemetry pipeline rather than disable sensors. Many malware-analysis systems adopt \emph{greedy} collection because they cannot predict which actions a program will take and thus lack clear collection thresholds. While endpoint products may block execution via static analysis, sandboxes are typically expected to record \emph{all} observable actions of a submitted binary; however, permissive logging can backfire. Since some filtering already exists, we focus on failures induced by \emph{data representation}.

Consider a sandboxed binary whose execution triggers process activity that is timestamped and linked via parent--child relationships, and may also produce artifacts such as memory dumps. This forms a tree, but interoperability and storage often use a nested JSON encoding. Arbitrary nesting, however, is not uniformly supported across libraries and backends.

Still, a binary that recursively spawns processes in a tree exhibits \emph{valid} behavior: modern OSes permit thousands of processes and sensors will log each creation event, forcing downstream components to serialize a potentially massive process tree. Common encodings (e.g., parent objects containing nested child objects) can exceed practical limits: many JSON libraries and document stores enforce nesting-depth thresholds (often $\sim$64--128 levels) and document-size caps, while dashboards may fail to render extreme hierarchies due to DOM growth or timeout constraints. In our experiments, recursive spawning readily crossed these limits, triggering serialization exceptions or truncation and rendering web-based UIs unresponsive. This gap between what sensors can collect and what backends can process creates exploitable failure modes across the pipeline.

To confirm that TCAs do not impede malware functionality, our payload executes a malicious action at a chosen depth (e.g., launching a shell), enabling systematic stress testing across commercial and open-source sandboxes and EDRs. The attack operates entirely in user mode, without code injection or interference with kernel hooks or sensor processes. Under the induced telemetry load, we observe dropped/incomplete logs, insertion errors, serialization failures, and dashboard instability; critically, the malicious action is often absent from reports, leading to missed detections and alerts. This evasion can be more damaging than a crash by falsely reassuring defenders that no threat is present.

\begin{figure}[th]
    \centering
    \includegraphics[width=\linewidth]{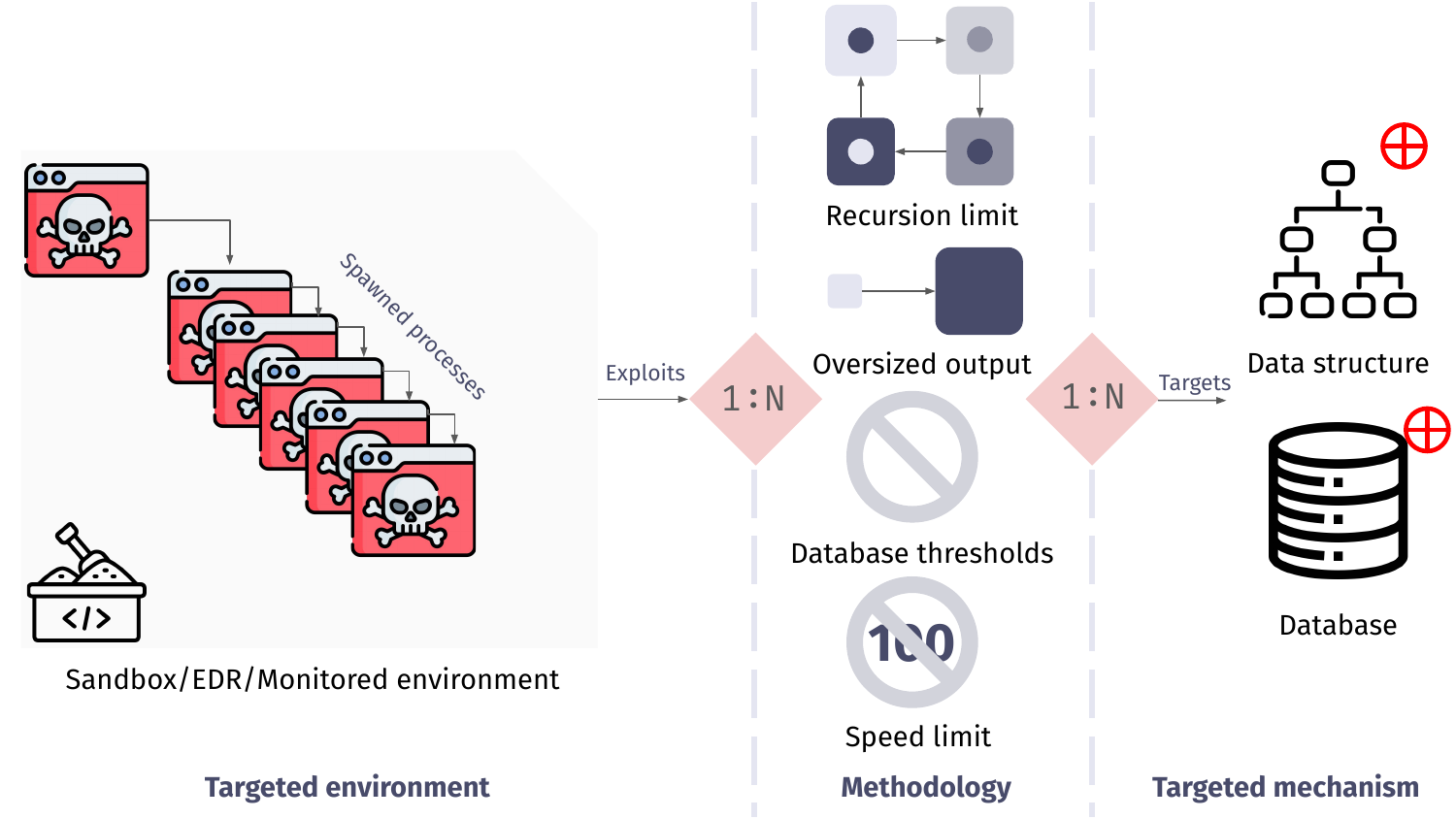}
    \caption{High-level overview of attack methodology}
    \label{fig:methodology}
\end{figure}

Our attack uses controlled recursion using a crafted binary that repeatedly launches an embedded copy of itself, increments a depth counter each iteration, and, at a predetermined maximum depth, executes a lightweight payload (in our tests, a PowerShell reverse shell) to confirm end-to-end execution. Depth is carried across processes via environment variables. The implementation evolved from a minimal C/C++ proof-of-concept into a hardened test binary with multiple stages of XOR-based encryption to reduce trivial static detection and ensure the relevant behavior manifests only at runtime.

\autoref{fig:methodology} summarizes the method's workflow and the primary failure modes exercised. On the left are the targeted environments (sandboxes and endpoint agents) in which a proof-of-concept binary was executed. The central column represents the operational limits we triggered in our experiments: \emph{recursion limit} refers to code-level or timeout protections that abort deep traversal; \emph{oversized output} captures large JSON or BSON blobs that exceed serializer or database limits; \emph{database thresholds} indicate document size and nesting caps or insert limits that cause errors; and \emph{speed limit} symbolizes UI rate caps that produce agent queue overflows or unresponsive dashboards. The panel to the right shows the telemetry mechanisms we targeted (process and call-argument data structures and persistent storage), where failures manifest as truncated or missing process trees, corrupted or absent behavioral reports, and UI stalls. Together, the elements illustrate how a simple recursive load can propagate through the telemetry pipeline and produce denial-of-analysis conditions.

\subsection{Threat model and assumptions} Our attacks require no elevated privileges, agent tampering, or vendor-specific vulnerabilities, and do not aim for persistence, exfiltration, privilege escalation, or cryptographic bypasses.
We detail our threat model and assumptions below. Thus, we list all the parameters that are relevant for our attack, what is controlled and by whom, but also what we expect to achieve.
\noindent\textbf{Execution context.} We assume that an adversary can store an unprivileged user-mode binary on either (i) an endpoint that is instrumented by sandboxing or EDR/DFIR tooling, or (ii) at a cloud solution, by uploading the binary through common user interfaces. The binary recursively spawns child processes up to a chosen depth and rate. Execution occurs under typical analysis conditions (e.g., detonation VMs, lab hosts, or enterprise-managed endpoints) where process creation is permitted, and environment variables propagate to child processes.

\noindent\textbf{Telemetry.} We assume that monitoring stacks capture process execution (e.g., via user-mode hooking and/or kernel/event providers), serialize these events into structured records (e.g., JSON/BSON or comparable encodings), and store them to a database-backed backend that exposes dashboards or APIs to analysts. Default configurations of such tools (e.g., document size caps, recursion/nesting limits, per-index write limits, dashboard rate caps, and analysis timeouts) are used as-is and not tuned to any specific workload. Also, the adversary cannot modify or disable sensors, hooks, or ETW/Sysmon providers; load kernel drivers; or gain elevated privileges, directly access or reconfigure serialization libraries, databases, or dashboards. Moreover, we assume the telemetry pipeline ingests data derived from untrusted program behavior (e.g., deep process trees, long command-line arguments, large argument arrays, and nested object graphs), and that serialization and storage components treat these data as valid inputs rather than as isolation boundaries.

\noindent\textbf{Resource bounds.} We assume that recursive spawners use a maximum depth chosen to avoid gross host-level denial of service (e.g., kernel scheduler collapse). We do not rely on exhausting OS resources (e.g., CPU, RAM, PIDs) as the primary failure mechanism. We confirmed this across all tested cases.

\noindent\textbf{Network and isolation.} According to our evaluation methodology, execution can occur in offline or network-restricted environments. A network is not required for the core effect of the presented methodology.

\noindent\textbf{Scope note (EDR architecture and applicability).}
EDR products fall into two architecturally distinct categories with respect to TCAs.
\emph{Prevention-first} EDRs (e.g., Kaspersky EDR, Microsoft Defender, WatchGuard EDPR) prioritize real-time blocking: once a process exhibits fork-bomb-like behavior, the agent terminates it before the telemetry pipeline is stressed. In these products the attack fails because execution is stopped, not because serialization or storage is robust; their resilience is therefore orthogonal to the pipeline vulnerabilities we study.
\emph{Telemetry-collector} EDRs (e.g., Wazuh) are architecturally closer to sandboxes and DFIR platforms. Their primary design goal is to capture and retain high-fidelity behavioral records for analyst review, not to make blocking decisions in real time. These products share the same unbounded-collection, bounded-serialization architecture that makes sandboxes vulnerable, and we demonstrate concrete DoA failures in both.
Our strongest and most general results therefore apply to platforms whose design goal is telemetry retention. That is, malware sandboxes, DFIR tools (e.g., Velociraptor), and telemetry-collector EDRs (e.g., Wazuh), where permitting execution and retaining telemetry are design requirements. We report outcomes for prevention-first EDRs separately as a boundary on applicability.
\subsection{Formal Model}
\begin{definition}[Telemetry pipeline]
We define a telemetry pipeline as a tuple
$\Pi = (\mathcal{A}, \mathcal{E}, \mathcal{S}, \mathcal{R})$
where the \emph{agent/collector} $\mathcal{A}$ observes system events
$e \in \Sigma$ and produces a stream
$X = (e_1, e_2, \dots)$; the \emph{encoder} $\mathcal{E}$ serializes batches
$B \subseteq X$ into objects $o = \mathcal{E}(B)$ belonging to an \emph{object space} $\mathcal{O}$ (e.g., JSON or BSON documents);
the \emph{store} $\mathcal{S}$ persists objects subject to capacity constraints;
and the \emph{renderer/API} $\mathcal{R}$ exposes persisted telemetry to analysts.
\end{definition}

\begin{definition}[Telemetry graph and depth]
    Execution induces a directed telemetry graph
$G = (V, E, D)$
where vertices $v \in V$ represent monitored entities, i.e., discrete observable objects generated or manipulated during execution (e.g., processes, threads, sockets), edges $E$ represent relations (e.g., spawned, communicated-with), and $D$ stores per-vertex/edge metadata (arguments, timestamps, command line).
For any directed graph $G$ with distinguished root vertex $r \in V$, we define its \emph{reachable depth} as
\[
d(G) \;=\; \max_{v \in V \text{ reachable from } r} \; \mathrm{dist}(r,v),
\]
where $\mathrm{dist}(r,v)$ is the length (number of edges) of a shortest directed path from $r$ to $v$. For a rooted tree, this coincides with the usual notion: the number of edges on the longest root-to-leaf path. When the root is clear from context, we write simply $d(G)$.
\end{definition}

\begin{definition}[Graph-to-document encoding and nesting depth]\label{def:graph-to-document-encoding}
Let $Enc_{\Pi}$ denote the concrete encoding used by a platform's telemetry pipeline $\Pi$ that maps a rooted telemetry graph $G$ to one or more serialized objects $o \in \mathcal{O}$.
Many systems encode process relations as nested objects or nested arrays (e.g., \autoref{lst:json-processes} in the Appendix).
We define the \emph{encoding nesting depth} $d_{\mathrm{enc}}(G)$ as the maximum bracket, object, or array nesting depth in the serialized representation produced by $Enc_{\Pi}(G)$.
\end{definition}

Note that $d_{\mathrm{enc}}(G)$ can equal the graph depth $d(G)$ for tree encodings, but can also differ if the platform flattens edges into tables, performs chunking, or uses multiple documents.
In our evaluation, we attribute nesting-limit failures to violations of bounds on $d_{\mathrm{enc}}(G)$, not necessarily on $d(G)$.

\begin{definition}[Admission predicates]
    Each \emph{pipeline stage}
    $Y \in \{ \mathcal{E}, \mathcal{S}, \mathcal{R} \}$, 
corresponding respectively to the encoder, store, and renderer components of the telemetry pipeline, implements a Boolean admission predicate
\[
\mathsf{admit}_Y : \mathcal{O} \times C_Y \rightarrow \{\mathsf{true}, \mathsf{false}\}
\]
which returns $\mathsf{true}$ iff object $o\in\mathcal{O}$ can be accepted under capacity constraints $C_Y$.
\end{definition}

\begin{definition}[Capacity model]
    Each pipeline stage $Y \in \{ \mathcal{E}, \mathcal{S}, \mathcal{R} \}$ has a capacity vector
    \begin{itemize}
        \item \textbf{Encoder limits:} $\mathcal{C}_{\mathcal{E}} = (d^{\mathsf{nest}}_{\max}, s^{\mathsf{obj}}_{\max}, r^{\mathsf{stack}}_{\max})$ for maximum nesting depth, per-object size, and recursion/stack depth.
        \item \textbf{Store limits:} $\mathcal{C}_S = (s^{\mathsf{doc}}_{\max}, d^{\mathsf{store}}_{\max}, q_{\max})$ for document size, nesting depth (e.g., BSON), and insert quota.
        \item \textbf{Renderer/UI limits:} $\mathcal{C}_R = (m^{mem}_{max}, n^{DOM}_{max}, e^{msg}_{max}, q^{buf}_{max}, d^{UI}_{max}, \mu_R)$,
capturing (in order) memory, UI element count, maximum message size, event-queue buffer capacity,
maximum renderable tree depth under the UI's DOM/memory constraints, and service rate.
    \end{itemize}
    A stage accepts an object $o$ (e.g., an encoded event, stored document, or rendered UI node) iff $\mathsf{admit}_Y(o, \mathcal{C}_Y) = \mathsf{true}$; otherwise it truncates or rejects the input.
\end{definition}
\begin{definition}[Rendered view]
Let $Render_{\Pi}$ denote the platform's renderer that maps stored/decoded telemetry into the analyst-visible UI view. We define
$\mathsf{Rendered}_{\Pi}(G) \subseteq V$
as the subset of vertices whose associated telemetry is ultimately presented to the analyst after encoding, storage, and rendering (including any truncation or drops).
\end{definition}

\begin{definition}[Stage-aware complexity]
Let $G=(V,E,D)$ be a telemetry graph rooted at $r$. Moreover, let 
\begin{enumerate*}[label=(\roman*)]
    \item $d(G)=\mathrm{depth}(G,r),$
    \item $S(G)=$ serialized size of the encoded object(s),
    \item $n(G)=|\mathsf{Rendered}_{\Pi}(G)|$, the number of analyst-visible entities (and, when rendered as a tree, a proxy for UI node count), and
    \item $\lambda(G)$ represents the rate at which telemetry events are generated, controlled by the adversary's spawn rate.
\end{enumerate*}
We collect these into a stage-aware complexity vector
\[
\vec{C}(G)=\bigl(d(G),\, S(G),\, n(G),\, \lambda(G)\bigr).
\]
Each pipeline stage $Y \in \{\mathcal{E}, \mathcal{S}, \mathcal{R}\}$ induces thresholds on the relevant components of $\vec{C}(G)$ via its capacity vector $\mathcal{C}_Y$. The first three components ($d$, $S$, $n$) govern static structural limits at the encoder, store, and renderer, respectively, while $\lambda$ governs dynamic throughput limits and is used to model queue-based failures (see \autoref{thm:burst})
\end{definition}

\begin{definition}[Visibility]
Let $\mathcal{M} \subseteq V$ denote nodes representing malicious activity. Moreover, let $\mathsf{Rendered}_{\Pi}(G) \subseteq V$ denote the subset of vertices whose associated telemetry successfully passes through all pipeline stages and is presented to the analyst. We define the visibility of the pipeline as
\[
\mathrm{Vis}_{\Pi}(G) =
\frac{| \mathcal{M} \cap \mathsf{Rendered}_{\Pi}(G) |}
     {| \mathcal{M} |}
\in [0,1],
\]   
where we assume $|\mathcal{M}| > 0$ (i.e., the execution includes at least one malicious action).
\end{definition}

\begin{definition}[Denial-of-analysis (DoA)]
   Given a policy threshold $\tau \in (0,1]$, a telemetry complexity attack (TCA)
\emph{succeeds} when
$\mathrm{Vis}_{\Pi}(G) < \tau,$
and the processes associated with vertices in $\mathcal{M}$ complete their intended computation (i.e., are not terminated by the security product). 
\end{definition}

\noindent\textbf{Adversary model}

The adversary $\mathcal{A}dv$ controls the \emph{shape} of the telemetry graph $G$ via their program behavior, without elevated privileges or tampering with sensors. We model the adversary's capability through a 
parameterized spawning function:

\begin{definition}[Spawning function]
Let $\mathsf{spawn}(b, d, k)$ denote an adversary-controlled process-creation routine that generates a $b$-ary tree of depth $d$, where each spawned process carries metadata of size $k$ bytes (e.g., command-line arguments). The adversary selects parameters $(b, d, k)$ subject to OS-imposed limits on concurrent processes and argument length.
\end{definition}

Thus, the adversary selects parameters:
$\mathsf{spawn}(b,d,k)$ to achieve a DoA in various settings.

\begin{definition}[Serialization overhead]
\label{def:overhead}
Let $\alpha > 0$ denote the \emph{structural serialization overhead}, that is, the fixed number of bytes added per node by the encoding format, independent of node metadata. This includes field delimiters, keys, brackets, and type markers. For JSON encoding of a process node, typical values are $\alpha \in [50, 200]$ bytes depending on the schema.
\end{definition}

The encoded object for a full $b$-ary tree of depth $d$ with per-node metadata of size $k$ has approximate serialized size
\[
S(b, d, k) \approx N(b, d) \cdot (\alpha + k),
\]
where $N(b, d) = \frac{b^{d+1} - 1}{b - 1}$ for $b > 1$, and $N(1, d) = d + 1$ for the unary (linear chain) case.
\begin{theorem}[Depth bound $\Rightarrow$ encoder/store/UI failure]
If the effective encoding nesting depth of $G$ exceeds either the encoder or store nesting caps, i.e.,
\[
d_{\mathrm{enc}}(G) > \min(d_{\max}^{\text{nest}},~ d_{\max}^{\text{store}}, d^{\text{UI}}_{\text{max}}),
\]
then for objects $o$ encoding the corresponding subtree,
\[
\mathsf{admit}_{\mathcal{E}}(o, C_{\mathcal{E}})=\mathrm{false}
\quad\text{or}\quad
\mathsf{admit}_{\mathcal{S}}(o, C_{\mathcal{S}})=\mathrm{false},
\quad\text{or}\]
\[\mathsf{admit}_{\mathcal{R}}(o,\mathsf{C}_R)=\mathsf{false}
\]
and telemetry associated with the pruned subtree becomes \emph{invisible}.
\end{theorem}
Note that $d^{\text{UI}}_{\text{max}}$ depends on browser memory limits and DOM complexity. 

\begin{theorem}[Size bound $\Rightarrow$ store failure]
Let $M = s_{\max}^{\text{doc}}$ be the per-document limit. If
$S(b,d,k) > M,$
the store rejects the document. Let $d^*$ denote the minimum depth guaranteed to trigger a size-bound failure. Then,
\[
d^* \ge \Bigl\lceil \log_b \!\Bigl(1+(b-1)\frac{M}{\alpha+k}\Bigr) \Bigr\rceil - 1.
\]
\end{theorem}

\begin{theorem}[Burst/queue bound $\Rightarrow$ renderer/API failure]\label{thm:burst}
Let $B = B_R$ and $\mu = \mu_R$ be the queue buffer capacity and service rate from the renderer capacity vector $\mathcal{C}_R$. We model the renderer as an $M/M/1/B$ queue with event arrival rate $\lambda = \lambda(G)$. If $\lambda > \mu$, buffer overflow occurs in expected time
\[
T_{\mathsf{overflow}} \approx \frac{B}{\lambda - \mu}.
\]
Additionally, if the renderer must materialize $N(b, d)$ DOM nodes and $N(b, d) > n^{DOM}_{max}$, or if the transport message size exceeds $e^{msg}_{max}$, the UI or API fails immediately.
\end{theorem}
Note that the $M/M/1/B$ model assumes Poisson arrivals. In practice, our recursive spawner generates deterministic, bursty arrivals. This model therefore provides an approximation of overflow timing; the actual overflow may occur faster under burst conditions. The DOM-size and message-size bounds hold independently of the arrival process.

\begin{definition}[Successful TCA]\label{def:successfulTCA}
Using the stage-aware complexity vector
\[
\vec{C}(G)=\bigl(d(G),\, S(G),\, n(G),\, \lambda(G)\bigr),
\]
a TCA succeeds iff at least one pipeline constraint is violated and visibility drops:
\[
\exists i:\ \vec{C}(G)_i > \vec{C}_{\max,i}
\quad\text{and}\quad
\mathrm{Vis}_{\Pi}(G) < \tau.
\]
\end{definition}

\begin{figure*}[t]
    \centering
    \resizebox{\linewidth}{!}{
        \begin{tikzpicture}[node distance=6mm and 1mm]

\tikzset{
  tcaFlow/.style={
    signal,
    signal from=west,
    signal to=east,
    align=center,
    minimum height=12mm,
    text depth=.25ex,
    font=\scriptsize,
  }
}

\node[tcaFlow, fill=col5, text=white] (poc) {Recursive\\ spawner (PoC)};
\node[tcaFlow, right=of poc, fill=col2] (A) {Agent/\\Collector $A$};
\node[tcaFlow, right=of A, fill=col6] (E) {Encoder $E$\\(JSON/BSON)};
\node[tcaFlow, right=of E, fill=col6] (S) {Store $S$\\ (DB/index)};
\node[tcaFlow, right=of S, fill=col6] (R) {Renderer/API\\$R$ (UI/query)};
\node[tcaFlow, right=of R, fill=col2] (ana) {Analyst/\\Verdict};

\node[font=\scriptsize, above=2mm of E,xshift=1.3cm]
  {Telemetry pipeline $\Pi=(A,E,S,R)$};

\node[tcaNoteL, above=8mm of E,xshift=-1cm] (knobs) {Adversary controls telemetry shape via program behavior:\\
$\textsf{spawn}(b,d,k)$ and event rate $\lambda$.\\
Stage-aware complexity: $\vec{C}(G)=(d,\;S(b,d,k),\;N(b,d),\;\lambda)$.};

\draw[tcaCallout] (knobs.south) -- (E.north);

\node[tcaNoteR, below=8mm of E,
  xshift=-2cm] (depth) {\textbf{Depth / nesting}\\
Trigger: $d>\min(d^{\max}_{\text{nest}},\,d^{\max}_{\text{store}})$\\
Effect: recursion errors or subtree truncation.};

\node[tcaNoteR, below=8mm of S,xshift=2cm] (size) {\textbf{Size / document limits}\\
Trigger: $S(b,d,k) > s^{\max}_{\text{doc}}$\\
Effect: rejected inserts, missing/empty reports.};

\node[tcaNoteR, above=8mm of R,xshift=1cm] (burst) {\textbf{Burst / UI + transport limits}\\
Trigger: $\lambda>\mu$ and/or message $>g_{\max}$ or DOM $>n^{\max}_{\text{DOM}}$\\
Effect: queue overflow, unresponsive dashboards.};

\draw[tcaCallout] (depth.north) -- (E.south);
\draw[tcaCallout] (size.north)  -- (S.south);
\draw[tcaCallout] (burst.south) -- (R.north);

\node[tcaFlow, right= of ana, fill=col5, text=white]
  (doa) {Denial-of-analysis\\
  $\mathrm{Vis}_{\Pi}(G) < \tau$\\while payload executes};

\end{tikzpicture}
    }
    \caption{A recursive process spawner induces a telemetry graph whose depth, size, or event rate exceeds implicit limits in the encoder, store, or renderer, yielding missing or unusable analyst-visible telemetry.}
    \label{fig:methodology2}
\end{figure*}

Therefore, we target implicit constraints of the underlying implementations for each telemetry component. More precisely:
First, we target serialization as implementations impose recursion, stack, and object size limits that can be triggered by nested or high-volume telemetry structures. Then, we target storage since backends impose document, request, or record limits that may reject inserts or truncate stored telemetry. For instance, when a platform uses MongoDB-like document stores, per-document size and nesting constraints can produce missing subtrees or empty reports. Finally, we target the rendering engines. Web UIs and APIs can fail due to oversized responses or when asked to materialize large hierarchical datasets; practical failures often manifest as timeouts, unresponsive dashboards, or incomplete report rendering.

Based on the above, we can assume a successful attack on a MongoDB-dependent solution as follows.
\begin{corollary}[MongoDB-like store limits]
\label{cor:mongodb}
Consider a deployment where $s^{\mathsf{doc}}_{\max} = 16\,\mathrm{MB}$ and $d^{\mathsf{store}}_{\max} = 100$ (following MongoDB's limits~\cite{mongo}). Let $\alpha$ be the structural serialization overhead. A sufficient condition to violate at least one store constraint is:
\[
d > 100 \quad \text{or} \quad d \geq \left\lceil \log_b\!\left(1 + \frac{(b-1) \cdot 16 \times 10^6}{\alpha + k}\right) \right\rceil - 1.
\]
\end{corollary}
Therefore, forcing a solution to exceed any of these thresholds individually, we expect an analysis failure (e.g., truncation, rejection, timeouts, UI hangs, or crashes).

Using the above formalisation, we can interpret \autoref{fig:methodology2} as a visual representation of Definition \ref{def:successfulTCA}. The recursive spawner controls the complexity vector $\vec{C}(G)$, and each failure mode corresponds to a violated capacity constraint in the encoder, store, or renderer. Moreover, our formal model enables us to estimate capacity parameters through documentation review and incremental probing.

\begin{table}[!ht]
\centering
\caption{Predicted vs. observed failure thresholds for platforms with known capacity parameters.
$\tau_{\text{pred}}$ and $\tau_{\text{obs}}$ denote the predicted and observed triggering thresholds
($d^*$ = chain depth; $N^*$ = total spawned processes).
Notation: \textsuperscript{\dag}MongoDB BSON nesting cap.
\textsuperscript{\ddag}gRPC transport cap; $\alpha$ calibrated from observed error (see text).
\textsuperscript{\S}Browser DOM node cap, from process-count stress test (\autoref{fig:cuckoo-proc-count}).
\textsuperscript{\P}Wazuh \texttt{analysisd} default event-queue capacity~\cite{WazuhDocs2023}.}
\label{tab:capacity-params}
\rowcolors{2}{}{gray!10}
\scriptsize
\begin{tabular}{llccc}
\toprule
\textbf{Platform} & \textbf{Binding limit} & \textbf{Thm.} & $\tau_{\text{pred}}$ & $\tau_{\text{obs}}$ \\
\midrule
CAPEv2      & MongoDB nest depth $= 100$\textsuperscript{\dag}        & 1 & $d^*=100$       & $d=100$             \\
Velociraptor & gRPC cap $= 4$\,MB, $\alpha=840$\,B\textsuperscript{\ddag} & 2 & $N^*=4993$   & $N=4500$--$6000$ \\
Cuckoo       & DOM nodes $\approx 5000$\textsuperscript{\S}           & 3 & $N^*=5000$   & $N\approx5000$   \\
Wazuh        & Queue $B=16384$, $e\approx7$/proc\textsuperscript{\P}  & 3 & $N^*\approx2341$ & $N\approx3000$ \\
\bottomrule
\end{tabular}
\end{table}
Using Theorems~1--3, we derive predicted failure thresholds for each platform where binding capacity parameters are publicly available or empirically observable. \autoref{tab:capacity-params} summarises results; we elaborate more in the Appendix.

\section{Experimental process and 
setup}
\label{sec:process}

\subsection{Setup}

As described in \autoref{sec:methodology}, we construct a binary that progressively peels encryption layers by spawning a decrypted copy of itself. Each execution maintains a process counter to determine the termination step; at that step, the binary opens a reverse shell to a host under our control. This creates deeply nested process structures (cf.\ \autoref{lst:json-processes} in the Appendix), intended to stress storage and processing paths and to expose truncation or handling failures. We introduced minor binary changes per target to probe solution-specific failure modes.

Experiments followed a progressive evaluation strategy. We began with open-source sandboxing platforms to iterate rapidly and observe internal failures, and then we moved to open-source EDR/DFIR tools and locally deployed trial instances of commercial products. Finally, we submitted sanitized samples to public cloud-based analysis services to validate behavior in online providers. Local testing was performed in isolated, offline laboratory networks using virtual machines to prevent telemetry leakage and to ensure safe handling of test artifacts. Each sandbox/EDR was installed with default or documented configurations; where supported, we varied timeouts, process-depth limits, memory-dump options, and report-compression settings to exercise distinct code paths. We repeated tests to ensure reproducibility and, when possible, cross-validated failures across independent environments. For each run, we monitored host resource usage (CPU, memory, disk I/O) to correlate system load with telemetry failures. Offline isolation also reduced vendor feedback effects (e.g., rapid signature updates or patches) that could otherwise suppress execution during the measurement window.

\subsection{Goals and evaluation criteria}
The primary objective of the experimental phase was to validate whether recursive process spawning could cause denial-of-analysis or otherwise impair the behavioral visibility of sandbox and endpoint detection systems. We designed the experiments to observe three main outcomes: \begin{enumerate*}[label=(\roman*)] \item denial or corruption of analysis reports, \item partial or total loss of behavioral telemetry, and \item degradation of analysis performance or stability leading to timeouts or crashes.\end{enumerate*}

Each local test environment, whether sandbox, EDR, or DFIR platform, was evaluated using identical input parameters, including recursion depth and number of spawned processes. The consistency of execution ensured that the differences in outcomes reflected genuine architectural or implementation differences between the platforms rather than sample variance. We considered a test successful if the sample prevented a complete report or caused pipeline instability; resilient if it fully logged recursive activity and correctly labeled behavior.

We considered a test successful if the sample prevented the behavioral engine from generating a complete or accurate report or caused instability in the telemetry or storage pipeline. Conversely, a platform was considered resilient if it fully recorded the recursive activity and correctly labeled the behavior as malicious without disruption.

\subsection{Test environments and controls}
\noindent\textbf{Methodology.}
We evaluated our approach across heterogeneous platforms to use distinct telemetry pipelines and persistence architectures. We first tested \emph{open-source sandboxes} (Cuckoo, CAPEv2) to analyze behavior logging and process-tree handling; their JSON-centric reporting exposed truncation, serialization, and database size constraints. We next tested \emph{open-source EDRs and monitoring tools}: Wazuh to examine agent-based monitoring under high-volume recursive process creation (e.g., truncated alerts, Sysmon event loss, memory pressure), and Velociraptor to assess forensic and live-response visibility under recursive load. We then deployed \emph{commercial trial and locally hosted} solutions (Bitdefender GravityZone, Kaspersky EDR, WatchGuard EDPR, Comodo Valkyrie) in offline virtual machines to measure detection resilience without external telemetry leakage. Finally, after offline validation, we submitted sanitized samples to \emph{online sandboxes} (VirusTotal, AnyRun, Hybrid-Analysis, Malwation Threat Zone, Recorded Future Triage, OPSWAT Filescan.io, Comodo Valkyrie) to compare behavior across cloud providers.

All systems were instrumented to capture system events, agent logs, exported reports, and backend errors. When available, we retained raw JSON/BSON dumps, database error logs, and UI traces for post-mortem analysis. Proof-of-concept artifacts and reproduction materials were handled under institutional disclosure policies and stored on isolated lab infrastructure.

\noindent\textbf{Evasion and sample variants.}
To reduce early static detection, our samples implement a lightweight multi-stage ``onion'' encryption scheme: each runtime stage decrypts only the subsequent stage, preventing the full payload and final command-and-control instruction from being observable via static inspection while preserving reproducibility across testbeds.
We generated multiple samples sharing identical core logic while varying payload, build architecture, and encryption depth. For payload, we used three variants: (1) dropping the EICAR test file to trigger anti-malware; (2) a default, widely used reverse-shell payload commonly flagged by anti-malware; and (3) a keylogger payload. We compiled each variant for both x86 and x64. To study depth effects, we constructed onion chains of length 1000, 2000, 3000, 4000, 5000, 6000, 7000, 8000, 9000, 10000, and 15000. Excluding auxiliary testing malware used to assess our methodology, each anti-malware solution was evaluated against 66 samples, yielding 792 total experiments reported in \autoref{sec:results}.

\begin{table*}[!th]
    \centering
    \scriptsize
    \rowcolors{2}{}{gray!10}
    \resizebox{\linewidth}{!}{
    \begin{tabular}{llllcc}
    \toprule
         \textbf{Environment}& \textbf{Availability}  &\textbf{Type}& \textbf{Version} &\textbf{Attack} & \textbf{CVE} \\
    \midrule
         Bitdefender GravityZone EDR & Proprietary & EDR & N/A & \myxmark &\\
         Bitdefender GravityZone Sandbox & Proprietary & Malware sandbox & N/A & \myunknown &\\
         Malwation Threat Zone & Proprietary & Malware sandbox & N/A &\myunknown &\\
         CAPEv2 & Open source & Malware sandbox & Commit 52e4b43 &\mycheck & CVE-2025-61301\\
         sandbox.pikker.ee  & Open source & Malware sandbox & Cuckoo v2.0.7 & \mycheck & Pending\\
         cuckoo.cert.ee  & Open source & Malware sandbox & Cuckoo v2.0.7& \mycheck & Pending\\
         Cuckoo 3  & Open source & Malware sandbox & 0.10.0-51-gad46ffe & \mycheck & Pending\\
         Kaspersky EDR & Proprietary & EDR & N/A &\myxmark &\\
         Microsoft Defender & Proprietary & EDR & N/A &\myxmark &\\
         AnyRun & Proprietary/Cloud & Malware sandbox & N/A & \myunknown &\\
         Recorded Future Triage & Proprietary/Cloud & Malware sandbox & N/A & \mycheck &  CVE-2025-61303\\
         Velociraptor & Open source & Digital forensic \& incident response & 0.74.1 &\mycheck  &\\
         VirusTotal & Proprietary/Cloud & Multiscan aggregator & N/A &\mycheck &\\
         WatchGuard EDPR & Proprietary & EDPR & N/A &\myxmark &\\
         Wazuh & Open source & EDR &
         4.12.0-rc1 &\mycheck &\\
         Hybrid-analysis& Proprietary/Cloud & Malware sandbox & N/A &\myunknown &\\
         OPSWAT MetaDefender& Proprietary/Cloud & Malware sandbox & N/A &\myunknown &\\
         Comodo Valkyrie & Proprietary/Cloud & Malware sandbox & N/A &\myunknown &\\
         \midrule
        orjson (library) & Open source & Serialization library & 3.11.4 & \mycheck & CVE-2025-67221 \\
    \bottomrule
    \end{tabular}
}
    \caption{Overview of our results across 18 evaluated platforms plus one dependency library (orjson, shown separately). Notation: \mycheck = achieved DoA/visibility loss while payload executed; \myxmark = blocked/fully logged; \myunknown = payload labeled as benign or undetected, but internal pipeline behavior could not be confirmed (black-box only).}
    \label{tbl:results}
\end{table*}

\section{Results}
\label{sec:results}
Due to variability in results across products, we detail the experimental results in the following paragraphs, categorizing them by attack success and product, and discussing the root causes for each product where appropriate. \autoref{tbl:results} summarizes the results and the severity of the failures observed.

For open-source platforms, we confirm failure modes using internal logs, code paths, and database artifacts. For proprietary or cloud systems, we often have only black-box evidence, such as truncated exported reports, API responses, UI behavior, and observable size or depth thresholds. In those cases, we describe causes as ``consistent with'' a given pipeline limit unless we can directly observe an explicit error, limit, or vendor confirmation. We retain raw artifacts (exported reports, error codes, and UI traces) to support these inferences.
Platforms marked \myunknown in \autoref{tbl:results} returned a benign verdict for our sample, but we were unable to determine whether this was due to a telemetry pipeline failure (true DoA) or simply because the product never triggered on the behavioral profile; in both cases, the malicious payload executed undetected.

\subsection{Wazuh}

Wazuh is a telemetry-collector EDR, rather than blocking on behavioral heuristics, it captures, indexes, and correlates endpoint events for analyst review. This architecture places it in the same vulnerability class as malware sandboxes; greedy collection feeding a bounded serialization and storage backend, and our results confirm this. Wazuh's pipeline is heavily dependent on JSON: Agents serialize endpoint events into JSON artifacts (e.g., \texttt{ossec-alerts.json}) and forward them for indexing and correlation. In our tests, adversarially structured telemetry caused those JSON streams and alert files to balloon (we observed \texttt{ossec-alerts.json} and related logs exceeding 100-200 MB), flooding the reporting and correlation pipeline. The result was not an obvious crash, but a silent loss of visibility; critical malicious activity was not logged or surfaced; many alerts were delayed or dropped, and numerous Windows Error Reporting files indicated local instability (stack-overflow exceptions) on the monitored hosts.

From an analyst's point of view, this is a denial-of-analysis. Wazuh continued to run but failed to produce useful, actionable telemetry, creating dangerous blind spots and false confidence. 
Therefore, the finding suggests a systemic risk for JSON-based EDR solutions, which can be exploited to hide part of their actions. 

\begin{figure}[t]
    \centering
    \begin{tikzpicture}
    \begin{axis}[
        width=0.98\columnwidth,
        height=0.52\columnwidth,
        xlabel={Number of spawned processes},
        ylabel={Outcome ID},
        xmin=0, xmax=10500,
        ymin=-0.5, ymax=5.5,
        ytick={0,1,2,3,4,5},
        yticklabels={0,1,2,3,4,5},
        grid=major,
        legend style={at={(0.5,1)},anchor=south,draw=none,fill=none},
        legend columns=3,
        line width=1.0pt,
        mark size=2.2pt,
        scaled x ticks=false
    ]

        \addplot+[mark=*] coordinates {
            (1,0)
            (1000,0)
            (2000,0)
            (3000,1)
            (4000,1)
            (5000,2)
            (5000,2) 
            (6000,2)
            (7000,4)
            (9000,4)
        };
        \addlegendentry{1/1}

        \addplot+[mark=square*] coordinates {
            (1000,0)
            (1000,0) 
            (2000,0)
            (2000,0) 
            (3000,1)
            (3000,1) 
            (4000,1)
            (4000,1) 
            (5000,2)
            (5000,2) 
            (6000,3)
            (7000,5)
            (8000,5)
            (10000,5)
        };
        \addlegendentry{1000/10}

        \addplot+[mark=triangle*] coordinates {
            (7000,4)
            (8000,4)
        };
        \addlegendentry{100/10 (flood / threads)}

    \end{axis}
    \end{tikzpicture}
    \caption{Agent-side stress outcomes versus process count under different flooding intensities. \\ (0) \emph{Flagged}, (1) \emph{Flagged} + \emph{Event queue flooded} \\
(2) \emph{NOT Flagged} + \emph{Event queue flooded} \\
(3) \emph{NOT Flagged} + \emph{Wazuh Error alert} \\
(4) \emph{NOT Flagged} + \emph{Not enough memory} \\
(5) same as (4) with anomaly marker in raw logs ($\ast$ or \texttt{/*})}
    \label{fig:agent-proc-flood}
    \vspace{-0.6cm}
\end{figure}

As seen in \autoref{fig:agent-proc-flood}, early runs are flagged despite successful shell execution; beyond $\sim$5000 processes, shell execution persists while detections disappear and the pipeline transitions to event-queue flooding and memory/resource errors. This experiment captures a qualitative phase change in defender visibility. At low to moderate loads (up to 2000 processes), the shell executes and the activity is flagged. As load increases, the agent/pipeline degrades: around 3000--6000 processes we observe explicit event-queue flooding and, importantly, loss of flagging even though shell execution continues. Past $\sim$7000 processes, the dominant failure mode becomes resource exhaustion (``not enough memory resources''), with additional anomalous cases (marked in the raw notes) under high-flood configurations, consistent with TCAs shifting detection from semantic analysis to capacity limits in telemetry handling.

\subsection{Velociraptor}
Velociraptor is a DFIR platform whose design goal is to collect and surface high-fidelity behavioral data for analyst review. As such, it does not make real-time blocking decisions; instead, it retains telemetry as comprehensively as possible, placing it in the same architectural vulnerability class, greedy collection against bounded serialization and transport, as confirmed in our results. The problem we observed was not a silent database rejection or a library exception, but rather an operational breakdown in Velociraptor's collection and rendering pipeline when asked to materialize very large process trees. Two related reproducible issues were identified during our testing.

First, the \texttt{Generic.System.Pstree} artifact attempts to return the entire process tree as a single JSON row. Under our recursive-load experiments, this row becomes extremely large and, during transmission or queuing, exhausts the client's memory buffers. The client process is then terminated by the local OS, producing the following observable behavior: the flow aborts with the message:
\begin{tcolorbox}[width=\columnwidth,colback={col1},boxsep=0pt]    
   \texttt{\small Flow not known - maybe the client crashed?}
\end{tcolorbox}
\noindent and the result folder for the collection contains only a tiny (4 KB) compressed artifact, indicating severe truncation of the expected output.
Velociraptor's engineering team confirmed the root cause; the artifact encoded the entire tree in a single JSON row. A patch was merged (PR \# 4454, Sep 17, 2025)~\cite{vpatch}
 with the commit message:
\begin{tcolorbox}[width=\columnwidth,colback={col1},boxsep=0pt]    
   \texttt{\small Bugfix: Limit number of entries emitted into process\_tracker\_tree()
. The Generic.System.Pstree artifact puts the entire tree into a single JSON row which can exceed buffer size in some cases. This causes problems in deeply nested trees.}
\end{tcolorbox}  

Second, even when the backend successfully receives the raw event stream, for example, when \texttt{Windows.Events.TrackProcesses} is configured to forward updates, the Velociraptor UI, and gRPC transport hit practical limits. In our tests, the raw sysmon-derived events did contain all spawned processes, but attempting to display or transmit that dataset produced the following gRPC error:
\begin{tcolorbox}[width=\columnwidth,colback={col1},boxsep=1pt]    
   \texttt{\small received message larger than max (8732115 vs. 4194304)}
\end{tcolorbox}

\noindent causing the interactive view to fail and the analyst to lose visibility. Therefore, although the events were captured, it was practically invisible to the operator, who had to perform low-level interactions to collect them.

\subsection{CAPEv2}
For CAPEv2, we used the commit \texttt{52e4b43} from 2025-05-17. CAPEv2 uses MongoDB for storage and relies heavily on JSON/BSON serialization in Python. MongoDB implies two distinct quotas. First, when storing analysis reports in  MongoDB, there is a 16\,MB BSON document size limit. Similarly, MongoDB enforces a maximum nesting depth of 100 levels. If reports exceed the 16\,MB quota or the nesting exceeds 100, the database rejects the insert, causing report generation to fail, regardless of whether the corresponding collector worked correctly.

Complementarily, CAPEv2 also relies on Python's \texttt{orjson} library to serialize results into JSON. However, the latter has its own limits regarding deeply nested or recursive structures. Therefore, this is another potential vulnerability condition, as properly crafted samples may cause the serializer to exceed Python's recursion limit, leading to serialization failures and aborted report generation. Note that through our investigation of CAPEv2, we identified a related vulnerability in orjson itself: the \texttt{orjson.dumps()} function does not enforce recursion limits when serializing deeply nested data structures, leading to stack exhaustion and crashes. This issue affects orjson versions through 3.11.4 and has been assigned CVE-67221~\cite{67221}. Since orjson is widely used in Python applications for high-performance JSON serialization~\cite{orjson_stats}, this vulnerability extends beyond CAPEv2 to any application that serializes 
untrusted nested data using orjson.

CAPEv2 had integrated mechanisms to mitigate the above, e.g., by progressively pruning large parts of the report by removing child nodes or entire process subtrees. However, this was proven insufficient when behavior complexity surpasses thresholds. Our tests resulted in a High-Severity (CVSS 7.5) denial-of-analysis vulnerability affecting \texttt{reporting/mongodb.py} and \texttt{reporting/jsondump.py} in CAPEv2  that allows attackers who can submit samples to cause incomplete or missing behavioral analysis reports by generating deeply nested or oversized behavior data that trigger MongoDB BSON limits or \texttt{orjson} recursion errors when the sample executes in the sandbox. Note that many systems depend on CAPEv2 dynamic analysis, e.g., VirusTotal uses it as one of its sandboxes.
This issue has been assigned CVE-61301~\cite{61301}. 

From a system perspective, system logs show MongoDB \texttt{OperationFailure} Code~15 and ``Recursion limit reached'' errors during JSON serialization. Moreover, final reports often contain no behavioral data or show ``failed reporting'', falsely making the sample appear benign or inactive. \autoref{tab:capev2_details} shows evidence collected across multiple analysis runs (report sizes, logged processes, and errors). As telemetry volume increases, runs transition from normal reporting to loss of analyst-visible behavior (`No behavior') and, ultimately, to reporting failure (missing/empty report) under recursion and storage constraints, see \autoref{tab:capev2_highlevel} in the Appendix.
\begin{table}[!ht]
\centering
\scriptsize
\rowcolors{2}{}{gray!10}
\begin{tabular}{@{}c c r r r l l@{}}
\toprule
\textbf{T} & \textbf{Prof} & \textbf{Dur} & \textbf{Proc} & \textbf{BSON} & \textbf{Report} & \textbf{Result} \\
\textbf{(s)} &  & \textbf{(s)} &  & \textbf{(files/KB)} & \textbf{(KB)} &  \\
\midrule
60  & L & 413 & 24 & 25/126 & 613  & OK \\
60  & B & 252 & 38 & 39/200 & 923  & OK \\
60  & H & 230 & 42 & 43/230 & 1054 & OK \\
100 & H & 330 & 61 & 62/318 & 1482 & OK \\
100 & B & 277 & 81 & 82/493 & 2032 & OK \\
120 & H & 340 & 72 & 73/383 & 1793 & OK \\
120 & B & --  & 88 & 89/475 & 2228 & OK \\
180 & H & 414 & 90 & 91/492 & 2356 & \textbf{No beh.} (MongoDB 15) \\
180 & B & 323 & 88 & 89/475 & 2228 & OK \\
200 & B & 413 & 119& 120/645& 3174 & \textbf{No beh.} (MongoDB 15) \\
200 & M & --  & -- & 144/726& --   & \textbf{Fail rpt.} (Rec.+Mongo) \\
300 & B & 513 & 146& 147/779& 3676 & \textbf{No beh.} (MongoDB 15) \\
300 & H & --  & -- & 167/892& --   & \textbf{Fail rpt.} (Rec.+Mongo) \\
\bottomrule
\end{tabular}
\vspace{2pt}
\begin{tabular}{@{}p{0.98\columnwidth}@{}}
\scriptsize
\textbf{Profiles (Prof):} \textbf{B}=baseline (no custom settings);\;
\textbf{L}=enforce-timeout;\;
\textbf{M}=full-proc-memory-dumps+import-reconstruction+enforce-timeout;\;
\textbf{H}=M + syscall. \\
\textbf{Abbrev:} OK=Behavior OK; No beh.=report indicates no behavior; Fail rpt.=failed reporting; Rec.=recursion limit reached.
\end{tabular}
\caption{CAPEv2: compact run-level evidence.}
\label{tab:capev2_details}
\end{table}
\autoref{tab:capev2_details} shows that ``\textit{No behavior}'' and ``\textit{Failed reporting}'' correlate with MongoDB insert failures (Code 15), and with recursion-limit errors under heavier logging profiles. Report size is a practical proxy for telemetry volume. For CAPEv2, failures cluster at higher volumes and longer timeouts above 180s (see \autoref{fig:capev2_outcomes_chart}); heavy profiles can fail reporting even when a timeout is fixed.

\begin{figure}[th]
\centering
\begin{tikzpicture}
\begin{axis}[
  width=\columnwidth,
  height=5.0cm,
  xlabel={Report size (KB)},
  ylabel={Timeout (s)},
  xmin=500, xmax=3800,
  ymin=40, ymax=320,
  ytick={60,100,120,180,200,300},
  ymajorgrids=true,
  xmajorgrids=false,
  tick label style={font=\scriptsize},
  label style={font=\scriptsize},
  legend style={
    font=\scriptsize,
    at={(0.5,1.28)},
    anchor=north,
    draw=none
  },
  legend columns=2,
]

\addplot+[only marks, mark=*, mark size=1.8pt] coordinates {
  (613,60) (923,60) (1054,60)
  (1482,100) (2032,100)
  (1793,120) (2228,120)
  (2228,180)
};
\addlegendentry{Behavior OK}

\addplot+[only marks, mark=triangle*, mark size=2.2pt] coordinates {
  (2356,180) (3174,200) (3676,300)
};
\addlegendentry{No behavior/Crash (MongoDB 15)}

\addplot+[only marks, mark=square*, mark size=2.2pt,mark options={fill=violet},color=violet] coordinates {
  (3775,200) (3775,300)
};
\addlegendentry{Failed reporting (Rec.+MongoDB)}

\end{axis}
\end{tikzpicture}
\caption{CAPEv2 outcomes vs. report growth.}
\label{fig:capev2_outcomes_chart}
\end{figure}

\vspace{-0.35in}
\subsection{Triage}
\label{subsec:triage}
Similar to our findings in CAPEv2, a critical (CVSS 9.8) denial-of-analysis vulnerability was found in Triage~\cite{triage}, a widely used commercial malware sandbox. Given that the platform is proprietary and the vendor did not disclose further details, we provide our own speculations. Because the API returns nested JSON process structures, the failure is consistent with depth/size limits in collection/serialization rather than pure UI rendering. This issue has been assigned CVE-61303~\cite{61303}.

Using the same concept as in CAPEv2, we submitted a sample to Triage for assessment. The sample generates a very high process volume and deeply nested activity. Using Windows 10 v2004 and Windows 10 LTSC 2021 images to test our sample, we brought the machine to a DoA condition, allowing a submitted sample to exhaust process-tracking and logging resources by recursively spawning large numbers of child processes. The result is truncated or missing telemetry; the PowerShell execution and reverse-shell stages that execute on the host are not recorded or reported. In fact, the initial behavioral output reported a score of 1/10, which marks it actually as benign, despite the sample executing PowerShell reverse-shell stages and other malicious actions on the host.

Reproducible public detonations are available on Triage~\cite{triage_sample}, which we used to validate the behavior across the provider's Windows images.
Based on the Triage API, we know that the behavior analysis report is a JSON, which, for the processes, uses a nested format. Therefore, we assume that the same process is followed when collecting the events, leading to a crash that truncates the output. We do not consider the issue to be in the presentation layer, as a score (1/10) is fetched and part of the process tree is reported. 

\begin{table}[!ht]
\centering
\scriptsize
\rowcolors{2}{}{gray!10}
\begin{tabular}{@{}l r r l@{}}
\toprule
\textbf{Phase} & \textbf{Timeout} & \textbf{Total} & \textbf{UI outcome} \\
\midrule
\textbf{OK}      & 30s-60s  & 2.9-7.7\,MB   & No lag \\
\textbf{Degraded} & 100s-120s & 20-22\,MB     & Lags (10-20\,s load) \\
\textbf{DoA}     & 180s      & 34\,MB         & Tab unresponsive \\
\textbf{DoA}     & 240s      & 35\,MB         & Unresponsive from start \\
\textbf{DoA}     & 300s      & 46\,MB         & Unresponsive from start \\
\textbf{Crash}   & 400s      & 51\,MB         & Tab crashed \\
\bottomrule
\end{tabular}
\caption{Cuckoo UI failure thresholds under telemetry growth.}
\label{tab:cuckoo-thresholds}
\end{table}

\subsection{\textit{Cuckoo}}
For \textit{Cuckoo}, our tests resulted in a denial-of-analysis in the behavior-logging and report-generation pipeline of \textit{CERT-EE}'s \textit{Cuckoo} deployments (including \texttt{\textit{cert-ee/cuckoo3}} commit \texttt{\textit{0.10.0-51-gad46ffe}} and \textit{Cuckoo} Sandbox \texttt{2.0.7} used by \textit{sandbox.pikker.ee} and \textit{cuckoo.cert.ee}) allows attackers who can submit samples for analysis to cause incomplete or missing behavioral reports by triggering unbounded recursive process creation that generates deeply nested or oversized behavior data. The excessive log volume can exceed the JSON or database serialization limits, leading to silent analysis failure or a frozen web interface.
We reproduced the issue on both public instances and locally cloned deployments. In the affected analyses, we regularly saw \texttt{process.json} and \texttt{registry.json} 
grow to 45-50 MB, so the underlying collectors were clearly logging a lot of activity; however, the final behavioral report was empty. Attempting to open those reports often caused the web UI to stall or the browser to freeze, and in offline clones, the large report files and UI instability prevented access to otherwise completed analyses. Representative runs and measured artifacts are summarized in \autoref{tab:cuckoo-thresholds}.
\begin{table}[!ht]
\centering
\scriptsize
\rowcolors{2}{}{gray!10}
\begin{tabular}{@{}c r r r r@{}}
\toprule
\textbf{Timeout} & \textbf{Total} & \textbf{registry.json} & \textbf{file.json} & \textbf{report.json} \\
\midrule
30s  & 2.9\,MB  & 2.6\,MB  & 183\,KB & 120\,KB \\
60s  & 7.7\,MB  & 7.1\,MB  & 471\,KB & 304\,KB \\
100s & 20\,MB   & 18\,MB   & 1.3\,MB & 823\,KB \\
120s & 22\,MB   & 20\,MB   & 1.4\,MB & 846\,KB \\
180s & 34\,MB   & 31\,MB   & 2.3\,MB & 1.5\,MB \\
240s & 35\,MB   & 32\,MB   & 2.3\,MB & 1.5\,MB \\
300s & 46\,MB   & 40\,MB   & 3.0\,MB & 1.8\,MB \\
400s & 51\,MB   & 46\,MB   & 3.4\,MB & 2.2\,MB \\
\bottomrule
\end{tabular}
\caption{Artifact composition driving the failure.}
\label{tab:cuckoo-composition}
\end{table}
\begin{figure}[!ht]
\centering
\begin{tikzpicture}
\begin{axis}[
  width=\columnwidth,
  height=5.0cm,
  xlabel={Timeout (s)},
  ylabel={Size (MB)},
  xmin=0, xmax=420,
  xtick={30,60,100,120,180,240,300,400},
  ymajorgrids=true,
  legend style={
    font=\scriptsize,
    at={(0.5,1.15)},
    anchor=north,
    draw=none
  },
  legend columns=4,
  tick label style={font=\scriptsize},
  label style={font=\scriptsize},
]

\addplot+[mark=*, thick] coordinates {
  (30,2.9) (60,7.7) (100,20.0) (120,22.0) (180,34.0) (240,35.0) (300,46.0) (400,51.0)
};
\addlegendentry{Total}

\addplot+[mark=none, thick] coordinates {
  (30,2.6) (60,7.1) (100,18.0) (120,20.0) (180,31.0) (240,32.0) (300,40.0) (400,46.0)
};
\addlegendentry{registry.json}

\addplot+[mark=none, thick,color=teal] coordinates {
  (30,0.179) (60,0.460) (100,1.3) (120,1.4) (180,2.3) (240,2.3) (300,3.0) (400,3.4)
};
\addlegendentry{file.json}

\addplot+[mark=none, thick] coordinates {
  (30,0.117) (60,0.297) (100,0.804) (120,0.826) (180,1.5) (240,1.5) (300,1.8) (400,2.2)
};
\addlegendentry{report.json}

\end{axis}
\end{tikzpicture}

\caption{Growth of artifacts as timeout increases.}
\label{fig:cuckoo-composition-lines}
\end{figure}
As shown in \autoref{tab:cuckoo-composition} and respective \autoref{fig:cuckoo-composition-lines}, \texttt{registry.json} dominates total size across all runs, making it the primary lever for UI overload; \texttt{file.json} and \texttt{report.json} grow steadily and contribute to the tipping points.

\section{Discussion}
\label{sec:discussion}
\subsection{Scalability and Generalization}
\label{sec:scalability}

To assess whether TCAs generalize beyond our primary proof-of-concept, we 
conducted additional experiments varying sample characteristics, recursion 
parameters, and execution environments. 

\noindent\textbf{Cross-architecture consistency.}
We compiled variants of our proof-of-concept for both x86 and x64 architectures and tested them on Windows 10, Windows 11, and Windows Server 2019. The recursion depth required to trigger failures did not vary across architectures and OS versions for a given platform, indicating that the vulnerabilities are not artifacts of specific binary formats or OS configurations.

\noindent\textbf{Reproducibility across runs.} Each configuration was executed a minimum of three times. For the local tests, since we had access to the backend, error codes were identical across runs at the same depth. Hence, we conclude that the observed failures are deterministic consequences of pipeline limits and failure points at the same recursion depths, so we report single values rather than distributions.

\noindent\textbf{Generalization to other recursive behaviors}
While our primary proof-of-concept uses process spawning, the underlying 
vulnerability (unbounded collection meeting bounded serialization) may 
manifest with other behaviors that generate deeply nested or high-volume telemetry, though this requires empirical validation. We identify three additional vectors that require further investigation.

First, we consider \emph{registry activity}. As sandboxes commonly serialize registry operations into hierarchical 
structures mirroring the registry's tree organization. So registry-focused TCAs may achieve comparable or greater 
impact with lower depth thresholds.

Then, we consider \emph{file system operations}, as deep directory traversal and creation generate telemetry that many 
platforms serialize with path nesting. Windows extended-length paths support up to 32,767 characters (\texttt{\textbackslash\textbackslash?\textbackslash} prefix), but serializers may not handle such depths gracefully.

Finally, we consider \emph{network connections}. High-frequency connection telemetry (e.g., thousands of short-lived 
sockets) stresses event queuing rather than nesting depth, potentially 
triggering the burst/queue failures modeled in ~\autoref{thm:burst}.

We leave empirical validation of these vectors to future work, noting that the architectural patterns we identified are shared across telemetry types within each platform.

\subsection{Root cause analysis and mitigation}
Based on our findings, several engineering measures can reduce DoA risk across monitoring platforms and mitigate TCAs. Monolithic, deeply nested JSON reports are fragile due to serializer depth/size limits; representing processes as explicit trees (or graph-backed models) is preferable. When nested encodings are unavoidable, producers should chunk large/deep structures and persist partial results to avoid full-report failure when limits are exceeded.

Telemetry should be validated at the agent/collector boundary against a strict schema and sanitized prior to ingestion. Any truncation must be explicitly marked with a standardized indicator to preserve analyst awareness and support recovery. For auditability, platforms should maintain an isolated, access-controlled forensic store that can accept raw artifacts when the primary pipeline rejects them.

Queue management and event collection are critical under heavy workloads. Spikes in event volume can trigger dropped events or collector crashes, and short-lived/recursive processes are especially prone to being missed, breaking parent--child linkage and producing ``orphan'' processes. Sandboxes can throttle bursts by delaying process handling, and can reduce volume by aggregating repetitive events (e.g., summarizing spawn sequences) and canonicalizing equivalent objects prior to transmission.

Finally, reporting paths must be robust to oversized datasets. Incremental rendering can prevent UI failures, while dashboards can prioritize key fields and expose truncated previews with clear indicators. Systems should estimate loadability given memory budgets and recursion/size thresholds, render lazily (e.g., virtualized trees, pagination), and avoid materializing large hierarchies in browser/server render threads. Streaming JSON/BSON parsers further reduce memory pressure by enabling partial consumption and checkpointing of long reports.

\subsection{Limitations}

While TCAs introduce new DoA conditions for anti-malware solutions, our study has limitations. First, although we tested multiple products, results may not generalize to all commercial/proprietary systems: architectural choices (telemetry pipelines, database backends, serialization libraries, data structures, and product goals) can change both applicability and outcomes. For example, EDRs may be less susceptible in practice because fast forking is an aggressive local pattern that can be mitigated by terminating the process after a small threshold without requiring host exposure.

Second, the failures we observed were largely concentrated in platforms built around Python and MongoDB and/or JSON/BSON serialization; systems using different languages, formats, or storage engines may exhibit different thresholds and error modes. This characterization is strongest for open-source systems where we could instrument internals end-to-end. By contrast, commercial solutions are black boxes: we only observe external behavior, so internal causes remain unknown. This also implies we may have missed relevant vulnerabilities (e.g., due to incorrect thresholds), and some outputs or missed detections could be artifacts of our attack rather than inherent weaknesses.

Finally, although we outline mitigations (e.g., back-pressure, incremental reporting, robust serialization), implementing and  evaluating these defenses across diverse products is beyond our scope.

\vspace{-0.4cm}

\section{Related work}
\vspace{-0.4cm}
\label{sec:related}
Dynamic analysis of binaries with and without the use of automated sandboxing is central to malware analysis. Prior work has shown various ways in which adversaries try to defeat these systems. Evasion of malware sandboxes usually aims at the identification of the sandbox environment through various means, such as timing attacks against virtualization, registry, network, or CPU artifact attacks, and human interaction detection ~\cite{10.1145/3150376.3150378}. For example, SandPrint showed that sandbox implementations may leak stable fingerprints that malware can detect and react to~\cite{yokoyama2016sandprint}.
Timing and resource channels remain practical. Embedding proof-of-work into malware exposes timing asymmetries between bare-metal and scaled cloud sandboxes, reducing the effectiveness of online analysis services~\cite{nappa2021powhow}. 

\paragraph{Attacks on EDRs}
While EDR systems offer far more comprehensive visibility than traditional antivirus solutions, they are not immune to evasion or abuse. Prior research and practitioner analyses have demonstrated multiple ways to bypass \cite{karantzas2021empirical,pogonin2022microsoft,junior2024hookchain} or even weaponize them \cite{safebreach,alachkar2025eviledr}. 

One of the main directions in this research line targets hooking mechanisms used by EDRs to intercept user and/or kernel-mode APIs. Several works exploit unhooking or direct system-call invocation to execute unmonitored code paths, typically by calling native Windows functions from \texttt{ntdll.dll} without passing through instrumented stubs \cite{apostolopoulos2021resurrecting,junior2024hookchain,10412066}. Because many EDRs rely on user-land hooks, attackers and defenders operate at the same privilege level, creating inherent race conditions that, in the end, favor attackers. Recent work, such as Cymulate's \emph{Blindside} technique, employed hardware breakpoints to selectively allow the loading of \texttt{ntdll.dll} while blocking other libraries~\cite{cymulate}.

Another common evasion vector uses legitimate binaries and scripts to execute malicious payloads, known as living-off-the-land tactics~\cite{lolbas}. 
Attackers may also abuse DLL sideloading, placing a malicious library next to a trusted executable so that it loads attacker-controlled code \cite{karantzas2021empirical,AQUARMOURY}. 
Control Panel (\texttt{.cpl}) files have likewise been repurposed to execute arbitrary payloads outside of standard inspection paths~\cite{ScareCrow,CPLResourceRunner}. 

The aforementioned methods have been widely used in the real world, as well as methods to terminate the EDR processes. Indeed, some attackers may resort to the use of vulnerable drivers to terminate the corresponding processes, as in the case of GhostEngine~\cite{GHOSTENGINE}. Moreover, following the Malware-as-a-Service paradigm, cybercriminals sell tools like Baphomet EDR Killer~\cite{EDRkillers}, which automate this process and disable EDRs.
Likewise, more advanced attackers have been reported manipulating Windows Defender Application Control (WDAC) policies to disable EDRs or whitelist malicious binaries \cite{WDACs}.  
More offensive strategies attempt to blind EDRs by blocking or corrupting their telemetry. For instance, tools such as \emph{EDRSilencer} use Windows Filtering Platform (WFP) rules to cut off network traffic to EDRs~\cite{EDRSilencer}, while \emph{EDR-Freeze}~\cite{EDRFreeze} exploits Windows Error Reporting to suspend the monitoring processes.

\paragraph{Attacks on malware sandboxes and security solutions}
Contrary to EDRs, malware sandboxes allow malware to execute, as their goal is to observe what the malware does when it is executed. To counter this, malware tries to hide as much information as possible. Therefore, malware needs to identify the environment in which it operates and determine whether to continue operating. For instance, since the malware sandbox is a virtualized environment, malware tries to determine whether it is being executed in a virtual machine. Side information, e.g., names of devices or running processes, hardware identifiers, screen resolution, available memory, or even available threads and processors, can easily give away the execution in a virtualized environment \cite{10.1145/3150376.3150378,184461,koutsokostas2021python}. Similarly, malware may try to find drivers, registry keys, or even folder names associated with virtual machines. Going a step further, malware may attempt to detect human-interaction elements to determine whether it is executed in a malware sandbox. For instance, the uptime, the lack of keyboard and mouse interaction, the absence of browser history, or recent file activity can be used to infer the environment. Therefore, many malware would perform these checks before execution of the actual payload. Moreover, this creates a highly fragile environment in which different configurations can yield radically different reports. 
Given the resource and time constraints, some malware may opt to delay their payload execution \cite{10.1145/2046707.2046740}. This forced malware sandboxes to manipulate time to trigger their execution faster. As a result of this arms race, malware tries to measure the timing of specific functions to determine whether the elapsed
time is aligned with the expected processing time. Evidently, time plays an important role in the dynamic analysis of malware \cite{KuchlerMHBB21}. For more on evasive malware, the interested reader may refer to \cite{GalloroPCCZ22}.

SlowFuzz automatically synthesizes inputs that trigger worst-case for several well-known algorithms, causing severe behavioral issues in widely deployed libraries \cite{petsios2017slowfuzz}. For what is relevant in this work, many various zip parsers used in antivirus software whose decompression was significantly delayed.
Other related work detected and repaired DoS-prone regex patterns, often used by cybersecurity solutions to detect strings or bytes associated with malicious activity (e.g., ReDoSHunter, RegexScalpel) and exposed vulnerabilities in non-backtracking matchers \cite{li2021redoshunter,turonova2022redos,li2022regexscalpel}. In content-inspection pipelines, adversarial inputs that induce super-linear or exponential matching can exhaust CPU budgets, delay processing, and cause packet or alert drops.

\section{Conclusions}
\label{sec:conclusions}
This work introduces TCAs, a novel method for inducing denial-of-analysis in modern anti-malware and telemetry monitoring systems by exploiting weaknesses in their data-handling and storage pipelines. Our findings are most directly applicable to systems that prioritize behavioral observation over real-time blocking, where permitting execution is a design requirement rather than a failure mode. We demonstrated that behavioral telemetry, not just sensors or hooks, can be an attack surface. Unlike traditional techniques that focus on hiding malicious behavior or evading detection, we directly target the internal telemetry infrastructure, demonstrating that unbound collection, serialization, and storage of monitoring data can itself be weaponized to disrupt analysis. Indeed, across 18 widely deployed products and cloud services, seven exhibited some form of DoA or behavioral invisibility. Our findings led to the assignment of four CVE identifiers, 
with one additional CVE pending for Cuckoo. Other vendors patched or 
reconfigured their software following coordinated disclosure.
 
Recursive workloads in experiments pushed JSON and BSON serialization to their limits, suggesting a need for more thorough stress testing of deeply nested and complex data structures to identify thresholds. Moreover, our results highlight the importance of viewing telemetry pipelines as potential attack surfaces and hardening them accordingly. Thus, in future work, we plan to explore fuzzing of telemetry structures and graph-based storage for behavioral data. Moreover, since Python and MongoDB were at the core of most observed denial-of-analysis conditions, it is essential to test other programming languages, serialization libraries, and database engines to check for similar weaknesses.  

\section*{Acknowledgment}
This work was supported by the European Commission under the Horizon Europe Programme, as part of the project SAFEHORIZON (GA 101168562). The content of this article does not reflect the official opinion of the European Union. Responsibility for the information and views expressed therein lies entirely with the authors.

\appendix

\section{Ethical Considerations}
\subsection{Stakeholders and potential impacts.}
Our work studies denial-of-analysis failure modes in telemetry pipelines used by defenders (security analysts, SOCs) and by vendors/operators of sandboxes, EDR, and DFIR tooling. The primary benefit is improved understanding of availability and robustness risks in widely deployed defensive systems, enabling mitigations that reduce blind spots for end users. A key risk is that the techniques could be misused by attackers to degrade defenders' visibility.

\subsection{Risk mitigation in experimentation.}
All local experiments were conducted in isolated, offline laboratory networks using virtual machines to prevent unintentional spread, external communication, or impact on third-party systems. When our test samples included network-capable behaviors (e.g., a reverse shell), they were configured to communicate only with a host under our control inside the isolated environment, and were not used against production systems. We did not target real user devices or collect personal data; logs, exported reports, and UI traces were retained only to support reproducibility and debugging of failure modes.

\begin{lstlisting}[language=json,caption={Nested JSON for spawned processes.},label={lst:json-processes}]
{"processes": {
  "id": "id1",
  "timestamp": 1761675797,
  "children": [
   {
    "id": "id2",
    "timestamp": 1761675798,
    "children": [
     {
      "id": "id3",
      "timestamp": 1761675797,
      "children": [
       {
        "id": "id4",
        "timestamp": 1761675799,
        "children": [...]
        ]
...]}
\end{lstlisting}

\subsection{Use of public/third-party services.}
For online sandbox validation, we submitted only sanitized samples intended to reproduce structural/telemetry stress behaviors without enabling deployment or abuse. Submissions were rate-limited and performed in a manner intended to minimize operational impact on service providers.

\subsection{Responsible and coordinated disclosure.}
All vulnerabilities discovered during this research were disclosed in accordance with coordinated and responsible disclosure practices. For each confirmed vulnerability, we followed a consistent process. \begin{enumerate*}[label=(\roman*)]  \item contact maintainers via their preferred security reporting channels within 48 hours of confirming exploitability; \item provide a 90-day disclosure window before public disclosure, with extensions upon request for lower-severity issues or issues requiring significant prerequisites, consistent with common disclosure policies (e.g., Project Zero~\cite{p0}); \item include technical details (affected versions, proof-of-concept, and suggested remediation when possible); and, where applicable, \item coordinate CVE assignment with maintainers who acknowledged the report and pursued remediation. When maintainers did not acknowledge the report, disputed that the issue constituted a vulnerability, or did not provide a remediation timeline, we proceeded with disclosure after the 90-day notification period, making reasonable efforts to minimize risk.\end{enumerate*} In some cases, issues without a published CVE were still disclosed to maintainers and resulted in patches or configuration hardening; in one case, a vendor did not consider the issue a vulnerability and granted permission to proceed with publication.

\textbf{Artifact release considerations.}
To reduce misuse risk prior to widespread deployment of defenses, we avoid releasing weaponized exploit chains or directly deployable binaries. Where artifacts are shared for review and reproducibility, we provide sanitized or structure-preserving variants sufficient to validate methodology and reproduce the reported resource/limit behaviors, and we document any omissions and their rationale.
  
\section{Open Science}
To facilitate reproduction and in accordance with the WOOT Security~'26 open-science policy, we provide the following artifacts:
\begin{itemize}
    \item Source code for generating the telemetry/malware-shaped inputs, parameterized by nesting depth and branching factor.
    \item Compiled binary sample of this code.
    \item Raw logss from some of the platforms.
    \item Links to proof-of-concept repositories corresponding to the disclosed vulnerabilities (with assigned CVEs), as referenced by the public CVE records.
\end{itemize}
All artifacts are available at \url{https://github.com/eGkritsis/PoC-Evidence-WOOT/}. For safety and responsible disclosure reasons, we do not include weaponized exploit chains or directly deployable binaries that could enable abuse prior to defenses; instead, we provide sanitized or structure-preserving variants sufficient to validate methodology and reproduce the reported behaviors, along with documentation describing any omitted components and the rationale for omission.

\section{Predicted vs observed failure thresholds}
\noindent\textbf{CAPEv2.} MongoDB enforces a BSON nesting depth cap of $d^{\text{store}}_{\text{max}} = 100$. Theorem~1 applies directly:
\begin{equation}
d^*_{\text{depth}} = d^{\text{store}}_{\text{max}} = 100.
\end{equation}
Observed failure at $d = 100$ confirms the prediction exactly.

\noindent\textbf{Velociraptor.} The gRPC transport cap is $s^{\text{msg}}_{\text{max}} = 4{,}194{,}304$\,B (from the logged error; see \autoref{sec:results}). The initial estimate of $\alpha = 150$\,B reflects bare structural overhead, but a full Sysmon process-creation record carries considerably more data. Back-calculating from the observed failure at $N \approx 5{,}000$ processes:
\[
\alpha_{\text{actual}} \approx \frac{4{,}194{,}304}{5{,}000} \approx 840\;\text{B/node.}
\]
Applying Theorem~2 with this calibrated value:
\[
N^* = \left\lfloor \frac{4{,}194{,}304}{840} \right\rfloor = 4{,}993,
\]
which falls squarely within the observed range of 4,500--6,000.

\noindent\textbf{Cuckoo.} The size-bound prediction from Theorem~2 ($N \approx 238{,}000$ for a 50\,MB limit with $\alpha = 210$) is never reached because the renderer fails first. \autoref{fig:cuckoo-proc-count} shows UI response time growing sharply and becoming unresponsive above $\sim5{,}000$ spawned processes, giving $n^{DOM}_{\text{max}} \approx 5{,}000$. Applying the DOM-node bound of Theorem~3:
\[
N^* = n^{DOM}_{\text{max}} \approx 5{,}000,
\]
consistent with the observed UI degradation threshold. This illustrates that renderer constraints can dominate well before serialization or storage limits.

\noindent\textbf{Wazuh.} Wazuh's failure is a throughput/queue failure rather than a depth or size failure, so Theorem~3 applies. The \texttt{analysisd} event queue has a documented default capacity of $B = 16{,}384$ events~\cite{WazuhDocs2023}. From the observed alert-file growth (100--200\,MB at $\sim7{,}000$ processes) and assuming an average alert record of $\sim2$\,KB, each spawned process generates approximately
\[
e \approx \frac{100 \times 10^6}{7{,}000 \times 2{,}048} \approx 7\;\text{events.}
\]
The queue overflows when the cumulative event count exceeds $B$:
\[
N^* = \left\lfloor \frac{B}{e} \right\rfloor = \left\lfloor \frac{16{,}384}{7} \right\rfloor \approx 2{,}341\;\text{processes.}
\]
Observed queue overflow begins at $\sim3{,}000$ processes, consistent with the estimate given approximation uncertainty in $e$.
These four cases together demonstrate that the formal model is \emph{predictive}: capacity parameters drawn from public documentation and logged error messages are sufficient to anticipate failure thresholds before running experiments, and the predictions align with the observed outcomes across different failure modes (depth, size, DOM, and queue).

\section{Undetected malicious behavior from malware sandboxes}
In the case of some malware sandboxes, e.g., Bitdefender GravityZone Sandbox and Threat Zone, even when there is a verdict, the solution does not detect what the malware was doing or report it to the analyst. In fact, in some cases, the solution might not have broken, e.g., Threat Zone, but the verdict was that the sample was benign.
\section{Failed attacks on EDRs}
Most EDRs we tested terminated the corresponding processes relatively early, whereas others completed the decryption and terminated the process once a connection to the attacking machine was established. Based on feedback from some vendors, they implemented this measure because their products detected an excessive number of spawned processes within a short time; therefore, they were flagged as fork bombs, and the corresponding processes were terminated. Clearly, EDRs do not have to record all interactions of a binary and wait for a final verdict. Once a binary triggers specific alerts, the EDR immediately kills the process, so our attack did not manage to succeed in most EDR solutions
\section{VirusTotal}
VirusTotal aggregates results from multiple antivirus engines and sandboxes, including CAPEv2 and other dynamic analysis backends. When we submitted our high-depth samples, several per-engine behavioral reports were absent or severely truncated. We attribute this to the CAPEv2 backend failure documented in \autoref{sec:results} since VirusTotal's aggregated behavioral view depends on contributing sandbox outputs, a DoA condition in CAPEv2 propagates silently to the VirusTotal report.

\section{UI/analysis degradation and process count stress test}

\begin{table}[!ht]
\centering
\small
\rowcolors{2}{}{gray!10}
\begin{tabular}{@{}l c l@{}}
\toprule
\textbf{Outcome} & \textbf{Timeout range} & \textbf{Observation} \\
\midrule
\textbf{Behavior OK} & 60-120s (most runs) & ID-A \\
\textbf{No behavior} & \(\ge\)180s (some runs) & ID-B \\
\textbf{Failed reporting} & 200s, 300s (heavy settings) & ID-C \\
\bottomrule
\end{tabular}
\caption{CAPEv2: failure thresholds at a glance.\\ \textit{ID-A: Reporting succeeds-report logged processes\\ID-B: Insertion fails (MongoDB Code 15) with execution\\ID-C: Encoder recursion limit + MongoDB Code 15; report missing}}
\label{tab:capev2_highlevel}
\end{table}

The CAPE pipeline exhibits non-linear failure modes under high process counts: at modest loads (e.g., 1 process), behavior is recorded, but at thousands of processes, the system often produces no behavioral output, and at 10000 processes the reporting stage can fail entirely. These outcomes were detected without tampering with sensors or instrumentation. Failure was induced by the volume of nested processes overwhelming downstream serialization and storage.

\begin{figure}[!ht]
    \centering
    \begin{tikzpicture}
    \begin{axis}[
        width=0.8\columnwidth,
        height=0.42\columnwidth,
        xlabel={Number of spawned processes},
        ylabel={Pipeline outcome},
        xmin=0, xmax=15500,
        ymin=-0.4, ymax=2.4,
        ytick={0,1,2},
        yticklabels={Failed reporting,No behavior,Behavior OK},
        grid=major,
        legend style={at={(0.5,1.02)},anchor=south,draw=none,fill=none},
        legend columns=2,
        scaled x ticks=false,
        xticklabel style={rotate=30}
    ]
        \addplot+[only marks, mark=*, mark size=1.8pt] coordinates {
            (1,2)
            (1000,1) (2000,1) (3000,1) (4000,1) (5000,1) (6000,1)
            (7000,2)
            (8000,1) (9000,1)
            (10000,0)
            (15000,1)
        };
        \addlegendentry{CAPE (default)}

        \addplot+[only marks, mark=square*, mark size=1.8pt] coordinates {
            (10000,1)
        };
        \addlegendentry{CAPE (with options)}
    \end{axis}
    \end{tikzpicture}
    \caption{CAPE process-tree stress results. Increasing the number of spawned processes causes missing behavior reports and, at higher loads, reporting failures (e.g., serialization/DB-layer errors), despite execution continuing in user mode.}
    \label{fig:cape-proc-count}
\end{figure}

In contrast to hard reporting failures, Cuckoo shows progressive degradation: UI latency grows from seconds to tens of seconds (and up to $\sim$80\,s around 9000 processes), and the process tree is frequently truncated while the page becomes unresponsive or slow. This behavior is consistent with telemetry pipelines that continue collecting events but cannot reliably transform and render large process graphs within practical time and resource limits.

\begin{figure}[!ht]
    \centering
    \begin{tikzpicture}
    \begin{axis}[
        width=0.95\columnwidth,
        height=0.42\columnwidth,
        xlabel={Number of spawned processes},
        ylabel={UI lag (s)},
        xmin=0, xmax=15500,
        ymin=0, ymax=85,
        ytick={0,20,40,60,80},
        grid=major,
        scaled x ticks=false,
        xticklabel style={rotate=30}
    ]
        \addplot+[mark=*, mark size=1.8pt] coordinates {
            (1,0)
            (1000,2)
            (2000,25)
            (3000,45)
            (4000,30)
            (5000,75)
            (6000,30)
            (7000,40)
            (8000,50)
            (9000,80)
            (10000,40)
            (15000,55)
        };
    \end{axis}
    \end{tikzpicture}
    \caption{Cuckoo process-count stress results. As the number of processes increases, the analysis UI degrades sharply, with long load times and tree truncation becoming common, indicating a progressive loss of analyst visibility before outright failure.}
    \label{fig:cuckoo-proc-count}
\end{figure}


\begin{thebibliography}{10}

  \bibitem{CrowdStrikeEDRDefinition}
  Anne Aarness.
  \newblock What is endpoint detection and response (edr)?
  \newblock CrowdStrike Cybersecurity 101, January 2025.
  \newblock Accessed Mar. 2, 2026.
  
  \bibitem{Afianian2018}
  Amir Afianian, Salman Niksefat, Babak Sadeghiyan, and David Baptiste.
  \newblock Malware dynamic analysis evasion techniques: {A} survey.
  \newblock {\em {ACM} Comput. Surv.}, 52(6):126:1--126:28, 2020.
  
  \bibitem{alachkar2025eviledr}
  Kotaiba Alachkar, Dirk Gaastra, Eduardo Barbaro, Michel van Eeten, and Yury
    Zhauniarovich.
  \newblock Eviledr: Repurposing {EDR} as an offensive tool.
  \newblock In Lujo Bauer and Giancarlo Pellegrino, editors, {\em 34th {USENIX}
    Security Symposium, {USENIX} Security 2025, Seattle, WA, USA, August 13-15,
    2025}, pages 587--605. {USENIX} Association, 2025.
  
  \bibitem{apostolopoulos2021resurrecting}
  Theodoros Apostolopoulos, Vasilios Katos, Kim{-}Kwang~Raymond Choo, and
    Constantinos Patsakis.
  \newblock Resurrecting anti-virtualization and anti-debugging: Unhooking your
    hooks.
  \newblock {\em Future Gener. Comput. Syst.}, 116:393--405, 2021.
  
  \bibitem{EDRSilencer}
  Chris Au.
  \newblock {EDRSilencer}.
  \newblock \url{https://github.com/netero1010/EDRSilencer}, 2023.
  \newblock Accessed Mar. 2, 2026.
  
  \bibitem{AVTEST2023}
  {AV-TEST Institute}.
  \newblock {Malware Statistics \& Trends Report}.
  \newblock \url{https://www.av-test.org/en/statistics/malware/}.
  \newblock Accessed Mar. 2, 2026.
  
  \bibitem{WDACs}
  Jonathan Beierle.
  \newblock {A Nightmare on EDR Street: WDAC's Revenge }.
  \newblock
    \url{https://beierle.win/2025-08-28-A-Nightmare-on-EDR-Street-WDACs-Revenge/},
    2025.
  \newblock Accessed Mar. 2, 2026.
  
  \bibitem{berlot2008dealing}
  Michele Berlot and Janche Sang.
  \newblock Dealing with process overload attacks in unix.
  \newblock {\em Inf. Sec. J.: A Global Perspective}, 17(1):33–44, January
    2008.
  
  \bibitem{AQUARMOURY}
  Soumyanil Biswas.
  \newblock {AQUARMOURY}.
  \newblock \url{https://github.com/reveng007/AQUARMOURY}, 2021.
  \newblock Accessed Mar. 2, 2026.
  
  \bibitem{GHOSTENGINE}
  Salim Bitam, Samir Bousseaden, Terrance DeJesus, and Andrew Pease.
  \newblock {Invisible miners: unveiling GHOSTENGINE's crypto mining operations}.
  \newblock
    \url{https://www.elastic.co/security-labs/invisible-miners-unveiling-ghostengine},
    2024.
  \newblock Accessed Mar. 2, 2026.
  
  \bibitem{CPLResourceRunner}
  Steve Borosh.
  \newblock {CPLResourceRunner}.
  \newblock \url{https://github.com/rvrsh3ll/CPLResourceRunner}, 2018.
  \newblock Accessed Mar. 2, 2026.
  
  \bibitem{SentinelOneBehavioralAI}
  Rick Bosworth.
  \newblock Decrypting sentinelone cloud detection | the behavioral ai engine in
    real-time cwpp.
  \newblock SentinelOne Blog, July 2025.
  \newblock Accessed Mar. 2, 2026.
  
  \bibitem{10.1145/3150376.3150378}
  Alexei Bulazel and B\"{u}lent Yener.
  \newblock A survey on automated dynamic malware analysis evasion and
    counter-evasion: Pc, mobile, and web.
  \newblock In {\em Proceedings of the 1st Reversing and Offensive-Oriented
    Trends Symposium}, ROOTS, New York, NY, USA, 2017. Association for Computing
    Machinery.
  
  \bibitem{cable_2024_13999026}
  Jack Cable.
  \newblock Ransomwhere: A crowdsourced ransomware payment dataset, October 2024.
  
  \bibitem{CiscoMerakiThreatGridIntegration}
  {Cisco}.
  \newblock Secure malware analytics ({Threat Grid}) integration.
  \newblock Cisco Meraki Documentation.
  \newblock Accessed Mar. 2, 2026.
  
  \bibitem{CiscoThreatGridDatasheet}
  {Cisco}.
  \newblock {Cisco Threat Grid Cloud: Data Sheet}.
  \newblock Technical report, Cisco, July 2021.
  \newblock Accessed Mar. 2, 2026.
  
  \bibitem{safebreach}
  Shmuel Cohen.
  \newblock The dark side of {EDR}: Repurpose {EDR} as an offensive tool.
  \newblock https://www.safebreach.com/blog/dark-side-of-edr-offensive-tool/,
    2024.
  
  \bibitem{crowdstrike}
  {CrowdStrike}.
  \newblock {CrowdStrike 2025 Global Threat Report}.
  \newblock \url{https://www.crowdstrike.com/en-us/global-threat-report/}, 2025.
  \newblock Accessed Mar. 2, 2026.
  
  \bibitem{EDRkillers}
  Roman Cuprik.
  \newblock {EDR killers get popular. Here is how to stop them}.
  \newblock
    \url{https://www.eset.com/blog/en/business-topics/threat-landscape/stop-edr-killers/},
    2025.
  \newblock Accessed Mar. 2, 2026.
  
  \bibitem{61301_poc}
  Constantinos~Patsakis Evgenios~Gkritsis, George~Stergiopoulos.
  \newblock {CVE-2025-61301}.
  \newblock \url{https://github.com/eGkritsis/CVE-2025-61301}, 2025.
  \newblock Accessed Mar. 2, 2026.
  
  \bibitem{61303_poc}
  Constantinos~Patsakis Evgenios~Gkritsis, George~Stergiopoulos.
  \newblock {CVE-61303}.
  \newblock \url{https://github.com/eGkritsis/CVE-2025-61303}, 2025.
  \newblock Accessed Mar. 2, 2026.
  
  \bibitem{67221_poc}
  Constantinos~Patsakis Evgenios~Gkritsis, George~Stergiopoulos.
  \newblock {CVE-2025-67221}.
  \newblock \url{https://github.com/kpatsakis/CVE-2025-67221}, 2026.
  \newblock Accessed Mar. 2, 2026.
  
  \bibitem{DBLP:conf/codaspy/FilicHMPU25}
  Mia Filic, Jonas Hofmann, Sam~A. Markelon, Kenneth~G. Paterson, and Anupama
    Unnikrishnan.
  \newblock Probabilistic data structures in the wild: {A} security analysis of
    redis.
  \newblock In James Joshi, Jaideep Vaidya, and Haya Schulmann, editors, {\em
    Proceedings of the Fifteenth {ACM} Conference on Data and Application
    Security and Privacy, {CODASPY} 2025, Pittsburgh, PA, USA, June 4-6, 2025},
    pages 167--178. {ACM}, 2025.
  
  \bibitem{FortinetSandboxingEbook}
  {Fortinet}.
  \newblock Unleashing the power of sandboxing: A crucial need for organizations.
  \newblock Technical report, Fortinet, November 2024.
  \newblock Accessed Mar. 2, 2026.
  
  \bibitem{GalloroPCCZ22}
  Nicola Galloro, Mario Polino, Michele Carminati, Andrea Continella, and Stefano
    Zanero.
  \newblock A systematical and longitudinal study of evasive behaviors in windows
    malware.
  \newblock {\em Comput. Secur.}, 113:102550, 2022.
  
  \bibitem{p0}
  {Google Project Zero}.
  \newblock Vulnerability disclosure policy, 2025.
  \newblock Accessed Mar. 2, 2026.
  
  \bibitem{ibm}
  {IBM}.
  \newblock Cost of a data breach report 2025.
  \newblock \url{https://www.ibm.com/reports/data-breach/}, 2025.
  \newblock Accessed Mar. 2, 2026.
  
  \bibitem{junior2024hookchain}
  Helvio~Carvalho Junior.
  \newblock Hookchain: {A} new perspective for bypassing {EDR} solutions.
  \newblock {\em CoRR}, abs/2404.16856, 2024.
  
  \bibitem{cymulate}
  Ilan Kalendarov.
  \newblock {EDR Evasion: A New Technique Using Hardware Breakpoints –
    Blindside}.
  \newblock
    \url{https://cymulate.com/blog/blindside-a-new-technique-for-edr-evasion-with-hardware-breakpoints/},
    2025.
  \newblock Accessed Mar. 2, 2026.
  
  \bibitem{karantzas2021empirical}
  George Karantzas and Constantinos Patsakis.
  \newblock An empirical assessment of endpoint detection and response systems
    against advanced persistent threats attack vectors.
  \newblock {\em Journal of Cybersecurity and Privacy}, 1(3):387--421, 2021.
  
  \bibitem{10.1145/2046707.2046740}
  Clemens Kolbitsch, Engin Kirda, and Christopher Kruegel.
  \newblock The power of procrastination: detection and mitigation of
    execution-stalling malicious code.
  \newblock In {\em Proceedings of the 18th ACM Conference on Computer and
    Communications Security}, CCS '11, page 285–296, New York, NY, USA, 2011.
    Association for Computing Machinery.
  
  \bibitem{koutsokostas2021python}
  Vasilios Koutsokostas and Constantinos Patsakis.
  \newblock Python and malware: Developing stealth and evasive malware without
    obfuscation.
  \newblock In Sabrina De~Capitani di~Vimercati and Pierangela Samarati, editors,
    {\em Proceedings of the 18th International Conference on Security and
    Cryptography, {SECRYPT} 2021, July 6-8, 2021}, pages 125--136. {SCITEPRESS},
    2021.
  
  \bibitem{KuchlerMHBB21}
  Alexander K{\"{u}}chler, Alessandro Mantovani, Yufei Han, Leyla Bilge, and
    Davide Balzarotti.
  \newblock Does every second count? time-based evolution of malware behavior in
    sandboxes.
  \newblock In {\em 28th Annual Network and Distributed System Security
    Symposium, {NDSS} 2021, virtually, February 21-25, 2021}. The Internet
    Society, 2021.
  
  \bibitem{10412066}
  Trevor~M. Lewis and Bhaskar~Prasad Rimal.
  \newblock Effects of removing user-land hooks in endpoint protection during
    attack experiments.
  \newblock {\em {IEEE} Access}, 12:15820--15844, 2024.
  
  \bibitem{li2021redoshunter}
  Yeting Li, Zixuan Chen, Jialun Cao, Zhiwu Xu, Qiancheng Peng, Haiming Chen,
    Liyuan Chen, and Shing{-}Chi Cheung.
  \newblock Redoshunter: {A} combined static and dynamic approach for regular
    expression dos detection.
  \newblock In Michael~D. Bailey and Rachel Greenstadt, editors, {\em 30th
    {USENIX} Security Symposium, {USENIX} Security 2021, August 11-13, 2021},
    pages 3847--3864. {USENIX} Association, 2021.
  
  \bibitem{li2022regexscalpel}
  Yeting Li, Yecheng Sun, Zhiwu Xu, Jialun Cao, Yuekang Li, Rongchen Li, Haiming
    Chen, Shing{-}Chi Cheung, Yang Liu, and Yang Xiao.
  \newblock Regexscalpel: Regular expression denial of service (redos) defense by
    localize-and-fix.
  \newblock In Kevin R.~B. Butler and Kurt Thomas, editors, {\em 31st {USENIX}
    Security Symposium, {USENIX} Security 2022, Boston, MA, USA, August 10-12,
    2022}, pages 4183--4200. {USENIX} Association, 2022.
  
  \bibitem{lolbas}
  {LOLBAS-Project}.
  \newblock {Living Off The Land Binaries, Scripts and Libraries}.
  \newblock \url{https://lolbas-project.github.io/}, 2026.
  \newblock Accessed Mar. 2, 2026.
  
  \bibitem{MSDefenderEDROverview}
  {Microsoft}.
  \newblock Overview of endpoint detection and response capabilities - microsoft
    defender for endpoint.
  \newblock Microsoft Learn, September 2025.
  \newblock Accessed Mar. 2, 2026.
  
  \bibitem{mongo}
  {MongoDB}.
  \newblock {MongoDB} limits and thresholds.
  \newblock
    \url{https://www.mongodb.com/docs/manual/reference/limits/?atlas-provider=aws&atlas-class=general},
    2024.
  \newblock Accessed Mar. 2, 2026.
  
  \bibitem{nappa2021powhow}
  Antonio Nappa, Panagiotis Papadopoulos, Matteo Varvello, Daniel~Aceituno Gomez,
    Juan Tapiador, and Andrea Lanzi.
  \newblock Pow-how: An enduring timing side-channel to evade online malware
    sandboxes.
  \newblock In Elisa Bertino, Haya Schulmann, and Michael Waidner, editors, {\em
    Computer Security - {ESORICS} 2021 - 26th European Symposium on Research in
    Computer Security, Darmstadt, Germany, October 4-8, 2021, Proceedings, Part
    {I}}, volume 12972 of {\em Lecture Notes in Computer Science}, pages 86--109.
    Springer, 2021.
  
  \bibitem{61301}
  {National Institute of Standards and Technology}.
  \newblock {CVE-2025-61301 Detail}.
  \newblock \url{https://nvd.nist.gov/vuln/detail/CVE-2025-61301}, 2025.
  \newblock Accessed Mar. 2, 2026.
  
  \bibitem{61303}
  {National Institute of Standards and Technology}.
  \newblock {CVE-2025-61303 Detail}.
  \newblock \url{https://nvd.nist.gov/vuln/detail/CVE-2025-61303}, 2025.
  \newblock Accessed Mar. 2, 2026.
  
  \bibitem{67221}
  {National Institute of Standards and Technology}.
  \newblock {CVE-2025-67221 Detail}.
  \newblock \url{https://nvd.nist.gov/vuln/detail/CVE-2025-67221}, 2025.
  \newblock Accessed Mar. 2, 2026.
  
  \bibitem{PANWAdvancedWildFireEnv}
  {Palo Alto Networks}.
  \newblock Advanced {WildFire} powered by {Precision AI}.
  \newblock Palo Alto Networks Documentation.
  \newblock Accessed Mar. 2, 2026.
  
  \bibitem{PANWCortexXDRWildFireReport}
  {Palo Alto Networks}.
  \newblock Review {WildFire} analysis details.
  \newblock Cortex XDR Documentation.
  \newblock Accessed Mar. 2, 2026.
  
  \bibitem{petsios2017slowfuzz}
  Theofilos Petsios, Jason Zhao, Angelos~D. Keromytis, and Suman Jana.
  \newblock Slowfuzz: Automated domain-independent detection of algorithmic
    complexity vulnerabilities.
  \newblock In Bhavani Thuraisingham, David Evans, Tal Malkin, and Dongyan Xu,
    editors, {\em Proceedings of the 2017 {ACM} {SIGSAC} Conference on Computer
    and Communications Security, {CCS} 2017, Dallas, TX, USA, October 30 -
    November 03, 2017}, pages 2155--2168. {ACM}, 2017.
  
  \bibitem{pogonin2022microsoft}
  Denis Pogonin and Igor Korkin.
  \newblock Microsoft defender will be defended: Memoryranger prevents blinding
    windows {AV}.
  \newblock {\em CoRR}, abs/2210.02821, 2022.
  
  \bibitem{orjson_stats}
  {PyPI Stats}.
  \newblock orjson, 2026.
  \newblock Accessed Mar. 2, 2026.
  
  \bibitem{8887385}
  Shawn Rasheed, Jens Dietrich, and Amjed Tahir.
  \newblock Laughter in the wild: {A} study into dos vulnerabilities in {YAML}
    libraries.
  \newblock In {\em 18th {IEEE} International Conference On Trust, Security And
    Privacy In Computing And Communications / 13th {IEEE} International
    Conference On Big Data Science And Engineering, TrustCom/BigDataSE 2019,
    Rotorua, New Zealand, August 5-8, 2019}, pages 342--349. {IEEE}, 2019.
  
  \bibitem{triage}
  {Recorded Future}.
  \newblock {Triage}.
  \newblock \url{https://tria.ge/}, 2025.
  \newblock Accessed Mar. 2, 2026.
  
  \bibitem{triage_sample}
  {Recorded Future}.
  \newblock {Triage sample}.
  \newblock \url{https://tria.ge/250822-qfystshk51}, 2025.
  \newblock Accessed Mar. 2, 2026.
  
  \bibitem{EDRFreeze}
  Zero Salarium.
  \newblock {EDR-Freeze}.
  \newblock \url{https://github.com/TwoSevenOneT/EDR-Freeze}, 2025.
  \newblock Accessed Mar. 2, 2026.
  
  \bibitem{184461}
  Hao Shi, Abdulla Alwabel, and Jelena Mirkovic.
  \newblock Cardinal pill testing of system virtual machines.
  \newblock In Kevin Fu and Jaeyeon Jung, editors, {\em Proceedings of the 23rd
    {USENIX} Security Symposium, San Diego, CA, USA, August 20-22, 2014}, pages
    271--285. {USENIX} Association, 2014.
  
  \bibitem{180226}
  Juraj Somorovsky, Andreas Mayer, J{\"{o}}rg Schwenk, Marco Kampmann, and Meiko
    Jensen.
  \newblock On breaking {SAML:} be whoever you want to be.
  \newblock In Tadayoshi Kohno, editor, {\em Proceedings of the 21th {USENIX}
    Security Symposium, Bellevue, WA, USA, August 8-10, 2012}, pages 397--412.
    {USENIX} Association, 2012.
  
  \bibitem{198467}
  Christopher Sp{\"{a}}th, Christian Mainka, Vladislav Mladenov, and J{\"{o}}rg
    Schwenk.
  \newblock Sok: {XML} parser vulnerabilities.
  \newblock In Natalie Silvanovich and Patrick Traynor, editors, {\em 10th
    {USENIX} Workshop on Offensive Technologies, {WOOT} 16, Austin, TX, USA,
    August 8-9, 2016}. {USENIX} Association, 2016.
  
  \bibitem{turonova2022redos}
  Lenka Turonov{\'{a}}, Luk{\'{a}}s Hol{\'{\i}}k, Ivan Homoliak, Ondrej
    Leng{\'{a}}l, Margus Veanes, and Tom{\'{a}}s Vojnar.
  \newblock Counting in regexes considered harmful: Exposing redos vulnerability
    of nonbacktracking matchers.
  \newblock In Kevin R.~B. Butler and Kurt Thomas, editors, {\em 31st {USENIX}
    Security Symposium, {USENIX} Security 2022, Boston, MA, USA, August 10-12,
    2022}, pages 4165--4182. {USENIX} Association, 2022.
  
  \bibitem{ScareCrow}
  Tylous.
  \newblock {ScareCrow}.
  \newblock \url{https://github.com/Tylous/ScareCrow}, 2023.
  \newblock Accessed Mar. 2, 2026.
  
  \bibitem{vpatch}
  {Velocidex}.
  \newblock {Bugfix: Limit number of entries emitted into
    process\_tracker\_tree() \#4454}.
  \newblock \url{https://github.com/Velocidex/velociraptor/pull/4454}, 2025.
  \newblock Accessed Mar. 2, 2026.
  
  \bibitem{WazuhDocs2023}
  {Wazuh, Inc.}
  \newblock {Wazuh Documentation --- Log Data Indexing and Storage}.
  \newblock
    \url{https://documentation.wazuh.com/current/getting-started/use-cases/log-analysis.html}.
  \newblock Accessed Mar. 2, 2026.
  
  \bibitem{yokoyama2016sandprint}
  Akira Yokoyama, Kou Ishii, Rui Tanabe, Yinmin Papa, Katsunari Yoshioka, Tsutomu
    Matsumoto, Takahiro Kasama, Daisuke Inoue, Michael Brengel, Michael Backes,
    and Christian Rossow.
  \newblock Sandprint: Fingerprinting malware sandboxes to provide intelligence
    for sandbox evasion.
  \newblock In Fabian Monrose, Marc Dacier, Gregory Blanc, and Joaqu{\'{\i}}n
    Garc{\'{\i}}a{-}Alfaro, editors, {\em Research in Attacks, Intrusions, and
    Defenses - 19th International Symposium, {RAID} 2016, Paris, France,
    September 19-21, 2016, Proceedings}, volume 9854 of {\em Lecture Notes in
    Computer Science}, pages 165--187. Springer, 2016.
  
  \end{thebibliography}
\end{document}